\documentclass[final]{ws-ijmpa}
\usepackage[super,compress]{cite}
\usepackage{graphicx}
\usepackage{epsfig}
\usepackage{amssymb}
\usepackage{url}
\pdfoutput=0

\usepackage{latexsym}
\usepackage{euscript}
\usepackage{times}
\usepackage{amsmath}
\usepackage{cite}

\newcommand{\beqs}{\begin{subequations}}
\newcommand{\eeqs}{\end{subequations}}
\newcommand{\ftn}{\footnotesize}

\newcommand{\ssz}{\scriptsize}

\newcommand{\Eref}[1]{Eq.~(\ref{#1})}
\newcommand{\Sref}[1]{Sec.~\ref{#1}}
\newcommand{\Fref}[1]{Fig.~\ref{#1}}
\newcommand{\Tref}[1]{Table~\ref{#1}}
\newcommand{\cref}[1]{Ref.~\citen{#1}}

\newcommand{\GeV}{{\mbox{\rm GeV}}}
\newcommand{\pb}{{\mbox{\rm pb}}}
\def\mcr{{\tt micrOMEGAs}}

\newcommand\vev[1]{\langle {#1} \rangle}

\newcommand{\staub}{\mbox{$\tilde \tau_2$}}
\newcommand{\bmm}{{\ensuremath{{\rm BR}\lf B_s\to \mu^+\mu^-\rg}}}
\newcommand{\bsg}{{\ensuremath{{\rm BR}\lf b\to s\gamma\rg}}}
\newcommand{\btn}{{\ensuremath{{\rm R}\lf B_u\to \tau\nu\rg}}}
\newcommand{\Dam}{{\ensuremath{\delta a_{\mu}}}}
\newcommand{\Omx}{{\ensuremath{\Omega_{\rm LSP} h^2}}}
\newcommand{\ssi}{{\ensuremath{\sigma^{\rm SI}_{\tilde\chi p}}}}
\newcommand{\ssd}{{\ensuremath{\sigma^{\rm SD}_{\tilde\chi p}}}}
\newcommand{\mx}{{\ensuremath{m_{\rm LSP}}}}
\newcommand{\Dst}{{\ensuremath{\Delta_{\tilde\tau_2}}}}
\newcommand{\Da}{{\ensuremath{\Delta_{A}}}}
\newcommand{\Dhp}{{\ensuremath{\Delta_{H}}}}

\newcommand{\mst}{{\ensuremath{m_{\tilde\tau_2}}}}
\newcommand{\Mg}{{\ensuremath{M_{1/2}}}}
\newcommand{\AMg}{{\ensuremath{A_0/M_{1/2}}}}
\newcommand{\xx}{{\ensuremath{\tilde\chi}}}
\newcommand{\tnb}{{\ensuremath{\tan\beta}}}
\newcommand{\sign}{{\ensuremath{\rm sign}}}
\newcommand{\Mgut}{\ensuremath{M_{\rm GUT}}}
\newcommand{\Ggut}{\ensuremath{G_{\rm PS}}}
\newcommand{\Gsm}{\ensuremath{G_{\rm SM}}}
\newcommand{\ldt}{\ensuremath{\lambda_{\bf 3}}}
\newcommand{\lds}{\ensuremath{\lambda_{\bf 1}}}
\newcommand{\ld}{\ensuremath{\lambda}}
\newcommand{\TeV}{\ensuremath{\rm TeV}}
\newcommand{\Ms}{\ensuremath{M_{\rm S}}}

\newcommand{\hh}{{\ensuremath{
I{\kern-2.6pt h}}}}
\newcommand{\bhh}{{\ensuremath{\bar{
I{\kern-2.6pt h}}}}}

\newcommand{\tr}{{\mbox{\sf\ssz T}}}
\newcommand{\Tr}{\mbox{\sf Tr}}

\newcommand{\ca}{\ensuremath{{\rm a}}}
\newcommand{\hd}{{\ensuremath{H_1}}}
\newcommand{\hu}{{\ensuremath{H_2}}}

\newcommand{\Fci}{\mbox{
$\llgm\bem q^c_{i1} \cr q^c_{i2} \cr q^c_{i3}\cr l^c_i\eem\rrgm$}}

\newcommand{\Fi}{\mbox{
$\llgm\bem q_{i1}&q_{i2}&q_{i3}&l_i\eem\rrgm$}}

\newcommand{\Hcb}{\mbox{
$\llgm\bem q^c_{H1} \cr  q^c_{H2} \cr q^c_{H3}\cr
l^c_H\eem\rrgm$}}

\newcommand{\Hca}{\mbox{
$\llgm\bem \bar q^c_{H1}&\bar q^c_{H2}&\bar q^c_{H3}&\bar
l^c_H\eem\rrgm$}}

\newcommand{\gmt}{\mbox{
$\llgm\bem \varepsilon_{\rm abc}g^c_{\rm c}  &\bar  g^c_\ca \cr
-\bar g^c_\ca & 0\eem\rrgm$}}

\newcommand{\dgmt}{\mbox{
$\llgm\bem \varepsilon_{\rm abc}\bar g^c_{\rm c}  &g^c_\ca \cr
-g^c_\ca & 0\eem\rrgm$}}

\newcommand{\bdhh}{{\ensuremath{\normalsize I{\kern-2.9pt H}}}}

\newcommand\lin[2]{\ensuremath{\llgm\bem #1& #2\eem\rrgm}}

\newcommand\stl[2]{\mbox{$\llgm\bem #1\cr #2\eem\rrgm$}}

\newcommand\lvc[3]{\ensuremath{\varepsilon_{{#1}{#2}{#3}}}}

\newcommand{\snH}{\ensuremath{\nu^c_H}}
\newcommand{\snHb}{\ensuremath{\bar\nu^c_H}}
\newcommand{\uH}{\ensuremath{u^c_H}}
\newcommand{\uHb}{\ensuremath{\bar u^c_H}}
\newcommand{\dH}{\ensuremath{d^c_H}}
\newcommand{\dHb}{\ensuremath{\bar d^c_H}}

\newcommand{\eHb}{\ensuremath{\bar e^c_H}}
\newcommand{\lH}{\ensuremath{\lambda_H}}
\newcommand{\lHb}{\ensuremath{\lambda_{\bar H}}}

\newcommand{\sni}{\ensuremath{\nu^c_i}}

\newcommand{\trc}{{\mbox{\sf\ssz T}}}

\def\openep{\boldsymbol{\varepsilon}}

\def\to{\rightarrow}

\def\lf{\left(}
\def\rg{\right)}

\def\llgm{\left\lgroup}
\def\rrgm{\right\rgroup}

\newcommand{\eem}{\end{matrix}}
\newcommand{\bem}{\begin{matrix}}
\def\beq{\begin{equation}}
\def\eeq{\end{equation}}
\def\bea{\begin{eqnarray}}
\def\eea{\end{eqnarray}}
\newcommand{\ec}{\end{center}}
\newcommand{\bec}{\begin{center}}

\renewenvironment{subequations}{%
\refstepcounter{equation}%
\setcounter{parentequation}{\value{equation}}%
  \setcounter{equation}{0}
  \def\theequation{\theparentequation{\sf\small\alph{equation}}}%
  \ignorespaces
}{%
  \setcounter{equation}{\value{parentequation}}%
  \ignorespacesafterend
}

\begin{document}
\markboth{N. Karagiannakis, G. Lazarides, and C. Pallis} {Cold
Dark Matter and Higgs Mass in the CMSSM with Generalized Yukawa
Quasi-Unification}

%
\catchline{}{}{}{}{}
%

\title{Cold Dark Matter and Higgs Mass in the
Constrained Minimal Supersymmetric Standard Model with Generalized
Yukawa Quasi-Unification}

\author{N. Karagiannakis}

\address{Physics Division, School of Technology,\\
Aristotle University of Thessaloniki, \\
54124 Thessaloniki, GREECE\\{\tt\small nikar@auth.gr}}

\author{G. Lazarides}

\address{Physics Division, School of Technology,\\
Aristotle University of Thessaloniki, \\
54124 Thessaloniki, GREECE\\{\tt\small lazaride@eng.auth.gr}}

\author{C. Pallis}

\address{Department of Physics, University of Cyprus, \\
P.O. Box 20537, CY-1678 Nicosia, CYPRUS\\
{\tt\small cpallis@ucy.ac.cy}}

\maketitle


\begin{abstract}
 The construction of specific supersymmetric grand unified
models based on the Pati-Salam gauge group and leading to a set of
Yukawa quasi-unification conditions which can allow an acceptable
$b$-quark mass within the constrained minimal supersymmetric
standard model with $\mu>0$ is briefly reviewed. Imposing
constraints from the cold dark matter abundance in the universe,
$B$ physics, and the mass $m_h$ of the lighter neutral CP-even
Higgs boson, we find that there is an allowed parameter space
with, approximately, $44\leq\tan\beta\leq52$, $-3\leq
A_0/M_{1/2}\leq0.1$, $122\leq m_h/\GeV\leq127$, and mass of the
lightest sparticle in the range $(0.75-1.43)~\TeV$. Such heavy
lightest sparticle masses can become consistent with the cold dark
matter requirements on the lightest sparticle relic density thanks
to neutralino-stau coannihilations which are enhanced due to
stau-antistau coannihilation to down type fermions via a
direct-channel exchange of the heavier neutral CP-even Higgs
boson. Restrictions on the model parameters by the muon anomalous
magnetic moment are also discussed.

\end{abstract}

\ccode{PACS numbers: 12.10.Kt, 12.60.Jv, 95.35.+d}

\maketitle

\section{Prologue}\label{sec:intro}

The constrained minimal supersymmetric standard model (CMSSM) is a
highly predictive version of the minimal supersymmetric standard
model (MSSM) based on universal boundary conditions \cite{Cmssm0,
Cmssm0a, Cmssm0b, Cmssm0c, Cmssm0d, Cmssm, Cmssma, Cmssmb,
Cmssmc}. It can be further restricted by being embedded in a
supersymmetric (SUSY) grand unified theory (GUT) with a gauge
group containing $SU(4)_{\rm c}$ and $SU(2)_R$. This can lead
\cite{pana, panaa} to `asymptotic' Yukawa unification (YU)
\cite{als, alsa}, i.e. the exact unification of the third
generation Yukawa coupling constants at the supersymmetric (SUSY)
GUT scale $M_{\rm GUT}$. In this scheme, we take the electroweak
Higgs superfields $H_1$, $H_2$ and the third family right handed
quark superfields $t^c$, $b^c$ to form $SU(2)_R$ doublets. As a
result, we obtain \cite{pana, panaa} the asymptotic Yukawa
coupling relation $h_t=h_b$ and, hence, large $\tan\beta\sim
m_{t}/m_{b}$. Furthermore, to get $h_b=h_{\tau}$ and, thus, the
asymptotic relation $m_{b}=m_{\tau}$, the third generation quark
and lepton $SU(2)_L$ doublets [singlets] $q_3$ and $l_3$ [$b^c$
and $\tau^c$] have to form a $SU(4)_{\rm c}$ {\bf 4}-plet
[${\bf\bar 4}$-plet], while the Higgs doublet $H_1$ which couples
to them has to be a $SU(4)_{\rm c}$ singlet. The simplest GUT
gauge group which contains both $SU(4)_{\rm c}$ and $SU(2)_R$ is
the Pati-Salam (PS) group $G_{\rm PS}=SU(4)_c\times SU(2)_L\times
SU(2)_R$ -- for YU within $SO(10)$, see Ref.~\citen{raby,
rabya,rabyb}.

Given the experimental values of the top-quark and tau-lepton
masses, the CMSSM supplemented by the assumption of YU (which
naturally restricts $\tan\beta\sim50$) yields unacceptable values
of the $b$-quark mass for both signs of the MSSM parameter $\mu$.
Moreover, the generation of sizable SUSY corrections \cite{copw,
copwa, copwb} to $m_b$ (about 20$\%$) drive it well beyond the
experimentally allowed region with the $\mu<0$ case being much
less disfavored. Despite this fact, we prefer to focus on the
$\mu>0$ case, since $\mu<0$ is strongly disfavored by the
constraint arising from the deviation $\delta a_\mu$ of the
measured value of the muon anomalous magnetic moment $a_\mu$ from
its predicted value $a^{\rm SM}_\mu$ in the standard model (SM).
Indeed, $\mu<0$ is defended \cite{g2davier} only at $3-\sigma$ by
the calculation of $a^{\rm SM}_\mu$ based on the $\tau$-decay
data, whereas there is a stronger and stronger tendency
\cite{Hagiwara, kinoshita} at present to prefer the
$e^+e^-$-annihilation data for the calculation of $a^{\rm
SM}_\mu$, which favor the $\mu>0$ regime. Note that, the results
of Ref.~\citen{Jen, Jena}, where it is claimed that the mismatch between
the $\tau$- and $e^+e^-$-based calculations is alleviated,
disfavor $\mu<0$ even more strongly.

The usual strategy to solve the aforementioned tension between
exact YU and fermion masses is the introduction of several kinds
of nonuniversalities in the scalar \cite{raby, rabya,rabyb,king,
kinga, kingb, baery, baerya, baeryb, baeryc} and/or gaugino
\cite{nath, natha, nathb} sector of MSSM with an approximate
preservation of YU. On the contrary, in Ref.~\citen{qcdm} -- see
also Refs.~\citen{muneg,nova,nova2,yqu,quasiShafi,pekino, pekinoa} --,
this problem is addressed in the context of the PS GUT model,
without the need of invoking departure from the CMSSM
universality. We prefer to sacrifice the exact YU in favor of the
universality hypothesis, since we consider this hypo\-thesis as
more economical, predictive, and easily accommodated within
conventional SUSY GUT models. Indeed, it is known -- cf. first
paper in \cref{shafi, shafia, shafia2, shafib} -- that possible violation of
universality, which could arise from D-term contributions if the
MSSM is embedded into the PS GUT model, does not occur provided
that the soft SUSY breaking scalar masses of the superheavy fields
which break the GUT gauge symmetry are assumed to be universal.

In the proposal of Ref.~\citen{qcdm}, the Higgs sector of the
simplest PS model \cite{leontaris,jean} is extended by including
an extra $SU(4)_{\rm c}$ nonsinglet Higgs superfield with Yukawa
couplings to the quarks and leptons. The Higgs $SU(2)_L$ doublets
in this superfield can naturally develop \cite{wetterich,
wettericha} subdominant vacuum expectation values (VEVs) and mix
with the main electroweak doublets, which are $SU(4)_{\rm c}$
singlets and form a $SU(2)_R$ doublet. The resulting electroweak
doublets $H_1$, $H_2$ break $SU(4)_{\rm c}$ and do not form a
$SU(2)_R$ doublet. Thus, YU is replaced by a set of \emph{Yukawa
quasi-unification conditions} (YQUCs) which depend on up to five
extra real parameters. The number of these parameters depends on
the representations used for the Higgs superfields which mix the
$SU(2)_L$ doublets in the $SU(4)_{\rm c}$ singlet and nonsinglet
Higgs bidoublets and some simplifying assumptions. These Higgs
superfields can either belong to a triplet or a singlet
representation of $SU(2)_R$. In the past \cite{qcdm, muneg, nova, nova2},
we have shown that the monoparametric YQUCs emerging from the
inclusion of one $SU(2)_R$-triplet [singlet] superfield could give
a SUSY model with correct fermion masses for $\mu>0$ [$\mu<0$].
However, only the model with $\mu>0$ could survive
\cite{nova,nova2,yqu,pekino, pekinoa} after imposing a set of
cosmological and phenomenological constraints. The same model can
also support new successful versions \cite{axilleas, axilleasa,
axilleasa2, axilleasb, axilleasc} of hybrid inflation, based solely on
renormalizable superpotential terms and avoiding overproduction of
monopoles \cite{monopole, monopolea, monopoleb, monopolec}. The
baryon asymmetry of the universe may be generated via nonthermal
leptogenesis \cite{nonthermallepto, nonthermalleptoa}.

However, the recently announced data -- most notably by the {\it
Large Hadron Collider} (LHC) -- on the mass of the SM-like Higgs
boson \cite{atlas,cms,cdf} as well as the branching ratio $\bmm$
of the process $B_s\to\mu^+\mu^-$ \cite{lhcb, lhcba} in
conjunction with {\it cold dark matter} (CDM) considerations
\cite{wmap} destroyed the successful picture above. To be more
specific, the upper bound from CDM considerations on the lightest
neutralino relic density, which is strongly reduced by
neutralino-stau coannihilations, yields a very stringent upper
bound on the mass of the lightest neutralino $m_\xx$, which is
incompatible with the lower bound on $m_\xx$ from the data
\cite{lhcb, lhcba,bsmumucombined} on $\bmm$. The main reason for
this negative result is that $\tan\beta$ remains large and, thus
-- see \Sref{sec:pheno} --, the SUSY contribution to $\bmm$ turns
out to be too large. To overcome this hurdle, we included in
\cref{gyqu} both $SU(2)_R$-triplet and singlet Higgs superfields.
This allows for a more general version of the YQUCs,  which now
depend on one real and two complex parameters. As a consequence,
the third generation Yukawa coupling constants are freed from the
stringent constraint $h_b/h_t+h_\tau/h_t=2$ obtained in the
monoparametric case and, thus, we can accommodate more general
values of the ratios $h_m/h_n$ with $m,n=t,b,\tau$, which are
expected, of course, to be of order unity for natural values of
the model parameters. Moreover, lower $\tnb$'s are allowed
reducing thereby the extracted $\bmm$ to a level compatible with
the CDM requirement. The allowed parameter space of the model is
then mainly determined by the interplay of the constraints from
$\bmm$ and CDM and the recently announced results of LHC on the
Higgs boson mass $m_h$.

In this review, we outline the construction of the SUSY GUT model
which can cause an adequate deviation from exact YU with sufficiently
low $\tnb$ so as the resulting CMSSM with $\mu>0$ to be consistent
with a number of astrophysical and experimental requirements. They
originate most notably from the data on $m_h$ and the $\bmm$ derived
by LHC and the nine-year fitting of the observations of the
\emph{Wilkinson microwave anisotropy probe} (WMAP) \cite{wmap} on
the CDM abundance. We show that the allowed parameter space of
the model is relatively wide, but the sparticle masses are too heavy
lying in the multi-TeV range. The latter signalizes a mild amount of
tuning as regards the achievement of the \emph{electroweak symmetry
breaking} (EWSB).

The construction of the model is briefly reviewed in Sec.~\ref{model}
and the resulting CMSSM is presented in Sec.~\ref{masspar}. The
parameter space of the CMSSM is restricted in Sec.~\ref{results}
taking into account a number of phenomenological and cosmological
requirements, which are exhibited in Secs.~\ref{sec:pheno} and
\ref{sec:cosmo} respectively. The deviation from exact YU is
estimated in Sec.~\ref{viol}. Finally, we summarize our conclusions
in Sec.~\ref{con}.

\section{The Pati-Salam Supersymmetric GUT Model}\label{model}

We outline below  -- in Sec.~\ref{gen} --  the salient features of
our model and then analyze the various parts of its superpotential
in Secs.~\ref{secWH}, \ref{secWm}, \ref{secWy}, and \ref{secWpq}.
Finally, we discuss the issue of the stability of the proton in
\Sref{sec:prot}.

\subsection{The General Set-up}\label{gen}

We focus on the SUSY PS GUT model which is described in detail in
Ref.~\citen{jean} -- see also Refs.~\citen{nova,nova2,nmH}. The
representations and the transformation properties under $G_{\rm
PS}$ of the various matter and Higgs superfields contained in the
model as well as their extra global charges (see below) are
included in Table~\ref{tab1}. The $i$th generation $(i=1,2,3)$
\emph{left handed} (LH) quark [lepton] superfields $u_{i\ca}$ and
$d_{i\ca}$ -- where $\ca=1,2,3$ is a color index -- [$e_i$ and
$\nu_i$] are accommodated in the superfields $F_i$. The LH
antiquark [antilepton] superfields $u^c_{i\ca}$ and $d_{i\ca}^c$
[$e^c_i$ and $\sni$] are arranged in the superfields $F^c_i$.
These superfields can be represented as
\bea \nonumber
&&F_i=\Fi\>\>\>\mbox{and}\>\>\>F^c_i=\Fci\>\>\>\mbox{with}\>\>\>\\
&& q_{i\ca}=\lin{d_{i\ca}}{-u_{i\ca}},
\>l_{i}=\lin{e_{i}}{-\nu_{i}},\>
q^c_{i\ca}=\stl{-u^c_{i\ca}}{d^c_{i\ca}}
,\>l^c_{i}=\stl{-\nu^c_{i}}{e^c_{i}}.
\label{Fg} \eea

\renewcommand{\arraystretch}{1.1}

\begin{table}[!t]
\tbl{\rm The representations and transformations under $G_{\rm
PS}$ as well as the extra global charges of the superfields of our
model ($U_{\rm c}\in SU(4)_{\rm c},~U_{L}\in SU(2)_{L},~U_{R}\in
SU(2)_{R}$ and $\tr~,\dagger$, and $\ast$ stand for the transpose,
the hermitian conjugate, and the complex conjugate of a matrix
respectively).} {\begin{tabular}{@{}cccccc@{}} \toprule {
Super-}&{ Represe-}&{ Trasfor-}&\multicolumn{3}{c}{ { Global} }
\\
\multicolumn{1}{c}{ fields}&{ ntations} &{ mations}
&\multicolumn{3}{c}{ { Charges}}
\\
\multicolumn{1}{c}{}&{ under $G_{\rm PS}$} &{ under $G_{\rm
PS}$}&{$R$} &{$PQ$} &{$\mathbb{Z}^{\rm mp}_2$}
\\\colrule
\multicolumn{6}{c}{ Matter Superfields}
\\ \colrule
{$F_i$} &{$({\bf 4, 2, 1})$}&$F_iU_L^{\dagger}U^\tr_{\rm c}$ &
$1/2$ & $-1$ &$1$
 \\
{$F^c_i$} & {$({\bf \bar 4, 1, 2})$}&$U_{\rm c}^\ast U_R^\ast
F^c_i$ &{ $1/2$ }&{$0$}&{$-1$}
\\ \colrule
\multicolumn{6}{c}{ Higgs Superfields}
\\ \colrule
{$H^c$} &{$({\bf \bar 4, 1, 2})$}& $U_{\rm c}^\ast U_R^\ast H^c$
&{$0$}&{$0$} & {$0$}
 \\
{$\bar H^c$}&$({\bf 4, 1, 2})$& $\bar{H}^cU^\tr_R U^\tr_{\rm
c}$&{$0$}&{$0$}&{$0$} \\
{$S$} & {$({\bf 1, 1, 1})$}&$S$ &$1$ &$0$ &$0$ \\
{$G$} & {$({\bf 6, 1, 1})$}&$U_{\rm c}GU^\tr_{\rm c}$ &$1$ &$0$
&$0$\\ \colrule
{$\hh$} & {$({\bf 1, 2, 2})$}&$U_L\hh U^\tr_R$ &$0$ &$1$ &$0$
 \\ \colrule
{$N$} &{$({\bf 1, 1, 1})$}& $N$ &{$1/2$}&{$-1$} & {$0$} \\
{$\bar N$}&$({\bf 1, 1, 1})$& $\bar N$&{$0$}&{$1$}&{$0$}
\\ \colrule
\multicolumn{6}{c}{ Extra Higgs Superfields}
\\ \colrule
$\hh^{\prime}$&{$({\bf 15, 2, 2})$} & $U_{\rm c}^\ast
U_L\hh^{\prime}U^\tr_RU^\tr_{\rm c}$ & $0$ & $1$ &$0$
 \\
 $\bhh^{\prime}$&{$({\bf 15, 2, 2})$} &
$U_{\rm c}U_L\bhh^{\prime}U^\tr_RU_{\rm c}^\dagger$ & $1$ & $-1$
&$0$
\\\colrule
$\phi$&$({\bf 15, 1, 3})$& $U_{\rm c}U_R\phi U_R^\dagger U_{\rm
c}^\dagger$ & $0$ & $0$ &$0$
\\
$\bar\phi$&{$({\bf 15, 1, 3})$} &$U_{\rm c}U_R\bar \phi
U_R^\dagger U_{\rm c}^\dagger$ & $1$ & $0$ &$0$
\\\colrule
$\phi'$&$({\bf 15, 1, 1})$& $U_{\rm c}\phi' U_{\rm c}^\dagger$ &
$0$ & $0$ &$0$
\\
$\bar\phi'$&{$({\bf 15, 1, 1})$} &$U_{\rm c}\bar \phi' U_{\rm
c}^\dagger$ & $1$ & $0$ &$0$
\\\botrule
\end{tabular}\label{tab1}}
\end{table}

\renewcommand{\arraystretch}{1.}

The gauge symmetry $G_{\rm PS}$ can be spontaneously broken down
to the SM gauge group $G_{\rm SM}$ through the VEVs which the
superfields
\bea \nonumber && H^c=\Hcb\>\>\>\mbox{and}\>\>\>\bar
H^c=\Hca\>\>\>\mbox{with}\>\>\>
\\ && q^c_{ H\ca}=\stl{ u^c_{H\ca}}{d^c_{H\ca}},\>
 l^c_{H}=\stl{\nu^c_{H}}{e^c_{H}},\> \bar q^c_{H\ca}=\lin{\bar
u^c_{H\ca}}{\bar d^c_{H\ca}},\> \bar
l^c_{H}=\lin{\bar \nu^c_{H}}{\bar e^c_{H}} \label{Hg} \eea
acquire in the direction $\nu^c_H$ and $\bar\nu^c_H$, respectively.
The model also contains a gauge singlet $S$, which triggers the
breaking of $G_{\rm PS}$, as well as a $SU(4)_{\rm c}$
{\bf 6}-plet $G$, which splits under $G_{SM}$ into a
$SU(3)_{\rm c}$ triplet $g_\ca^c$ and antitriplet $\bar{g}^c_\ca$,
which give \cite{leontaris} superheavy masses to $d^c_{H\ca}$ and
$\bar{d}^c_{H\ca}$. In particular, $G$ can be represented by an
antisymmetric $4\times4$ matrix
\beq G=\gmt\>\Rightarrow\>\bar G=\dgmt, \eeq
where $\bar G$ is the dual tensor of $G$ defined by $\bar
G_{IJ}=\varepsilon_{IJKL}G_{KL}$ and transforms under
$SU(4)_{\rm c}$ as $U_{\rm c}^*\bar G U_{\rm c}^\dagger$. Here,
$\varepsilon_{IJKL}$  [$\lvc{\ca}{\rm b}{\rm c}$] is the
well-known antisymmetric tensor acting on the
$SU(4)_{\rm c}$  [$SU(3)_{\rm c}$] indices with
$\varepsilon_{1234}=1$  [$\lvc{1}{2}{3}=1$]. The
symmetries of the model allow the presence of quartic
(nonrenormalizable) superpotential couplings of $\bar{H}^c$ to
$F^c_i$, which generate intermediate-scale masses for the
\emph{right handed neutrinos} $\sni$ and, thus, masses for the
light neutrinos $\nu_i$ via the seesaw mechanism.

In addition to $G_{\rm PS}$, the model possesses two global $U(1)$
symmetries, namely a Peccei-Quinn (PQ) \cite{pq, pqa, pqb} and a R
symmetry, as well as a discrete $Z_2^{\rm mp}$ symmetry (`matter
parity') under which $F$, $F^c$ change sign. Note that global
continuous symmetries such as our PQ and R symmetry can
effectively arise \cite{laz1} from the rich discrete symmetry
groups encountered in many compactified string theories -- see
e.g. Ref.~\citen{laz2, laz2a}.

In the simplest realization of this model \cite{nova,nova2,leontaris},
the electroweak doublets $H_1, H_2$ are exclusively contained in
the bidoublet superfield $\hh$, which can be written as
\beq \hh=\llgm\bem\hh_2&\hh_1\eem\rrgm, \eeq
and so the model predicts YU at $M_{\rm GUT}$ -- note that
$M_{\rm GUT}$ is determined by the requirement of the unification
of the gauge coupling constants. In order to allow for a sizable
violation of YU, we extend the model by including three extra
pairs of Higgs superfields $\hh',\bhh'$, $\phi,\bar\phi$, and
$\phi',\bar\phi'$, where the barred superfields are included in
order to give superheavy masses to the unbarred superfields. These
extra Higgs superfields together with their transformation
properties and charges under the global symmetries of the model
are also included in Table~\ref{tab1}. The two new Higgs superfields
$\hh^{\prime}$ and $\bhh^\prime$ with
\beq \hh^\prime=\llgm\bem
\hh^\prime_2&\hh^\prime_1\eem\rrgm~~\mbox{and}
~~\bhh^\prime=\llgm\bem\bhh^\prime_2&\bhh^\prime_1\eem\rrgm
\label{hs}\eeq
belong to the ({\bf 15,2,2}) representation of $SU(4)_{\rm c}$,
which is the only representation besides ({\bf 1,2,2}) that can
couple to the fermions. On the other hand, $\phi$ and $\phi'$
acquire superheavy VEVs of order $M_{\rm GUT}$ after the breaking
of $G_{\rm PS}$ to $G_{\rm SM}$. Their couplings with
$\bhh^{\prime}$ and $\hh$ naturally generate a ${SU(2)}_R$- and
${SU(4)}_{\rm c}$-violating mixing of the $SU(2)_L$ doublets in
$\hh$ and $\hh^{\prime}$ leading, thereby, to a sizable violation
of YU.

More explicitly, the superpotential $W$ of our model naturally
splits into four parts
\beq W=W_{\rm H}+W_{\rm M}+W_{\rm Y}+W_{\rm PQ},
\label{Wtotal}\eeq
which are specified, in turn, in the following Secs.~\ref{secWH},
\ref{secWm}, \ref{secWy}, and \ref{secWpq}.

\subsection{\boldmath The Spontaneous Breaking of $\Ggut$
to $\Gsm$}\label{secWH}

The part of $W$ in \Eref{Wtotal} which is relevant for the
breaking of $\Ggut$ to $\Gsm$ is given by
\bea W_{\rm H}&=&\kappa S\lf H^c\bar{H}^c-M^2\rg-S\lf \beta
\phi^2+\beta' \phi^{\prime2}\rg +\lf
\lambda\bar{\phi}+\lambda'\bar{\phi}'\rg H^c\bar{H}^c\nonumber \\
&& + m\phi\bar{\phi}+m'\phi'\bar{\phi}', \label{superpotential}
\eea
where the mass parameters $M,~m$, and $m'$ are of order $M_{\rm
GUT}$ and $\kappa$, $\beta$, $\beta'$, $\lambda$, and $\lambda'$
are dimensionless complex parameters. Note that by field redefitions
we can set $M,~m,~m',~\kappa,~\lambda$, and $~\lambda'$ to be real
and positive. For simplicity, we also take $\beta>0$ and $\beta'>0$
(the parameters are normalized so that they correspond to the
couplings between the SM singlet components of the superfields).

\par
The scalar potential obtained from $W_{\rm H}$ is given by
\begin{eqnarray}
V_{\rm
H}&=&\left\vert\kappa(H^c\bar{H}^c-M^2)-\beta\phi^2-\beta'\phi^{\prime2}
\right\vert^2+\left\vert\kappa S+\lambda\bar{\phi}
+\lambda'\bar{\phi}'\right\vert^2\left(\vert
H^c\vert^2+\vert\bar{H}^c \vert^2\right) \nonumber \\
&&+\left\vert m\phi+\lambda H^c \bar{H}^c\right\vert^2+\left\vert
m'\phi'+\lambda' H^c \bar{H}^c\right\vert^2 \nonumber\\
&& +\left\vert 2\beta S\phi-m\bar{\phi} \right\vert^2 +\left\vert
2\beta' S\phi'-m'\bar{\phi}' \right\vert^2+\ {\rm D-terms},
\label{potential}
\end{eqnarray}
where the complex scalar fields which belong to the SM singlet
components of the superfields are denoted by the same symbols as
the corresponding superfields. Vanishing of the D-terms yields
$\bar{H}^c\,^{*}=e^{i\vartheta}H^c$ ($H^c$, $\bar{H}^c$ lie in the
$\nu^c_H$, $\bar{\nu}^c_H$ direction). We restrict ourselves to
the direction with $\vartheta=0$, which contains the SUSY vacua
(see below). Performing appropriate R and gauge transformations,
we bring $H^c$, $\bar{H}^c$, and $S$ to the positive real axis.

From the potential in \Eref{potential}, we find that the SUSY
vacuum lies at
\beqs\beq
\label{vacuum}\vev{H^c\bar{H}^c}=v^2_0,~\vev{S}=\vev{\bar{\phi}}=\vev{\bar{\phi}'}=0
\eeq and \beq\vev{\phi}=v_\phi\left(T_{\rm
c}^{15},1,\frac{\sigma_3}{\sqrt{2}}
\right),~\vev{\phi'}=v'_\phi\left(T_{\rm
c}^{15},1,\frac{\sigma_{0}}{\sqrt{2}} \right),\label{vacuum3} \eeq
where \bea \lf{\frac{v_0}{
M}}\rg^2=\frac{1}{2\xi}\left(1-\sqrt{1-4\xi}\right), ~v_\phi=-{\ld
\frac{v_0^2}{ m}},~~{v'_\phi} =-{\ld' \frac{v_0^2}{ m'}}
\label{vacuum2} \eea with \beq
\xi={\frac{M^2}{\kappa}}\lf{\frac{\beta\lambda^2}{
m^2}}+{\frac{\beta'\lambda^{\prime2}}{m^{\prime2}}}\rg<1/4.\eeq
\eeqs The structure of $\vev{\phi}$ and $\vev{\phi'}$ with respect
to (w.r.t.) $G_{\rm PS}$ is shown in Eq.~(\ref{vacuum3}), where
\beqs \bea T^{15}_{\rm c}= \frac{1}{2\sqrt{3}}\>{\sf
diag}\lf1,1,1,-3\rg,~\sigma_{3}={\sf diag}\lf1,-1\rg,
~~\mbox{and}~~\sigma_{0}={\sf diag}\lf1,1\rg.\eea\eeqs

\subsection{Mass Terms}\label{secWm}

The part of $W$ in \Eref{Wtotal} which gives masses to the various
components of the superfields naturally splits into three parts
\beq W_{\rm M}=W_{G}+W_{\rm RHN}+W_{\rm mix} \label{WMeq}\eeq
out of which the first one is responsible for the generation of
superheavy masses for the superfields $\dH$ and $\dHb$:
\bea  \nonumber W_{G}&=&\lH H^{c\trc} G \openep H^c + \lHb
\bar{H}^{c} \bar G \openep \bar{H}^{c\trc}\\ &=&-2\lH\lf \nu^c_H
d^c_H -e^c_H u^c_H
\rg\bar{g}^c+ 2\lH u^c_H d^c_H g^c\nonumber\\
&&- 2\lHb \lf\bar{\nu}^c_H\bar{d}^c_H- \bar{e}^c_H\bar{u}^c_H\rg
g^c + 2\lHb\bar{u}^c_H \bar{d}^c_H \bar{g}^c,\label{WGeq}\eea
where the color indices have been suppressed and $\openep$ is the
$2\times 2$ antisymmetric matrix with $\openep_{12}=1$. Let us note,
in passing, that the combination of two [three] color-charged
objects in a term involves a contraction of the color indices with
the symmetric [antisymmetric] invariant tensor $\delta_{\ca\rm
b}~[\lvc{\ca}{\rm b}{\rm c}]$, e.g. $\uH\uHb=\delta_{\ca\rm b}
u_{H\ca}^c \bar u_{H\rm b}^c~[u^c_H d^c_H g^c=\lvc{\ca}{\rm b}{\rm
c} u^c_{H\ca} d^c_{H\rm b}g_{\rm c}^c]$. Given that $H^c$ and
$\bar H^c$ in \Eref{vacuum} acquire their VEVs along the direction of
$\snH$ and $\snHb$ respectively, it is obvious from \Eref{WGeq} that
$g^c$ and $\bar{g}^c$ pair with $\dHb$ and $\dH$ respectively and
acquire superheavy masses of order $M_{\rm GUT}$.

The second part $W_{\rm RHN}$ of $W_{\rm M}$ in \Eref{WMeq} provides
intermediate scale Majorana masses for $\sni$ as follows:
\begin{eqnarray} W_{\rm RHN}=\ld_{ij\nu^c}\bar{H}^cF_i^c\bar{H}^c
F_j^c/\Ms~~~~~~~~~~~~~~~~~~~~~~~~~~~~~~~~~~~~ \nonumber \\
=\ld_{ij\nu^c}(\eHb e_i^c+\dHb d_i^c-\snHb\nu_i^c-\uHb u_i^c)
(\eHb e_j^c+\dHb d_j^c-\snHb\nu_j^c-\uHb u_j^c)/\Ms, \label{WRHNeq}
\end{eqnarray}
where $\Ms\simeq 5\cdot 10^{17}~{\rm GeV}$ is the string scale.
Therefore,  the $\sni$'s acquire Majorana masses of order
$M_{\rm GUT}^2/\Ms\sim 10^{10}-10^{14}~{\rm GeV}$ depending on the
magnitude of the coupling constants $\ld_{ij\nu^c}$.

The last part of $W_{\rm M}$ in \Eref{WMeq}, which is responsible for the
mixing of the $SU(2)_L$ doublets in $\hh$ and $\hh'$, is a sum of $G_{\rm PS}$
invariants with the traces taken w.r.t. the $SU(4)_{\rm{c}}$ and $SU(2)_L$
indices:
\beq W_{\rm mix}=M_\hh \Tr\left(\bhh^{\prime}\openep
\hh^{\prime\tr}\openep\right)+\ldt\Tr\left(\bhh^{\prime}
\openep\phi \hh^\tr \openep \right) +\lds\Tr\left(
\bhh^{\prime}\openep\phi' \hh^\tr \openep\right).\label{Wm}\eeq
Here the mass parameter $M_\hh$ is of order $M_{\rm GUT}$ (made
real and positive by field rephasing) and $\ldt$, $\lds$ are
dimensionless complex coupling constants. Note that the two last
terms in the \emph{right hand side} (RHS) of \Eref{Wm} overshadow
the corresponding ones from the nonrenormalizable
$SU(2)_R$-triplet and singlet couplings originating from the
symbolic coupling $\bar{H}^cH^c\bhh^{\prime}\hh$ (see
Ref.~\citen{qcdm}).

Replacing $\phi$ and $\phi^\prime$ by their VEVs in
Eq.~(\ref{vacuum3}) and expanding the superfields in Eq.~(\ref{hs})
as linear combinations of the fifteen generators $T^a$ of
$SU(4)_{\rm c}$ normalized so as $\Tr(T^aT^b)=\delta^{ab}$
and denoting the colorless components of the superfields by the
superfield symbol, we can easily establish the following
identities:
\beqs\bea && \Tr\left(\bhh^{\prime} \openep \hh^{\prime\tr}
\openep\right)=\bhh^{\prime\tr}_1\openep \hh^\prime_2 +
\hh^{\prime\tr}_1\openep \bhh^\prime_2+\cdots,\label{idnta}
\\ &&
\Tr\left(\bhh^\prime\openep\phi\hh^\tr\right)=\frac{v_\phi}{\sqrt{2}}
\Tr\left(\bhh^{\prime}\openep\sigma_{3}\hh^\tr
\openep\right)=\left(\bhh^{\prime\tr}_1\openep \hh_2 -
\hh^\tr_1\openep\bhh^\prime_2\right),
\\ &&
\Tr\left(\bhh^\prime\openep\phi^\prime\hh^\tr\right)=\frac{v^\prime_\phi}
{\sqrt{2}}
\Tr\left(\bhh^{\prime}\openep\sigma_{0}\hh^\tr
\openep\right)=\left(\bhh^{\prime\tr}_1\openep \hh_2 +
\hh^\tr_1\openep\bhh^\prime_2\right),
\label{idnt}\eea\eeqs
where the ellipsis includes color nonsinglet components of the
superfields. Upon substitution of the above formulas in the RHS of
\Eref{Wm}, we obtain the mass terms
\beq W_{\rm mix}=M_\hh\bar
\hh^{\prime\tr}_1\openep\left(\hh^{\prime}_2+
\alpha_2\hh_2\right)\\
+M_\hh\left(\hh^{\prime \tr}_1+\alpha_1
\hh^\tr_1\right)\openep\bhh^{\prime}_2+\cdots,
\label{superheavy}\eeq
where the complex dimensionless parameters $\alpha_{1}$ and
$\alpha_{2}$ are given by
\beqs\bea\label{alphas1}
\alpha_{1}&=&{\frac{1}{\sqrt{2}M_\hh}}\lf-\lambda_{\bf
3}v_\phi+\lambda_{\bf 1}v'_\phi\rg,
\\ \alpha_{2}&=&{\frac{1}{\sqrt{2}M_\hh}}\lf\lambda_{\bf
3}v_\phi+\lambda_{\bf 1}v'_\phi\rg\cdot \label{alphas2} \eea\eeqs

It is obvious from Eq.~(\ref{superheavy}) that we obtain the
following two pairs of superheavy doublets with mass $M_\hh$
\beq\label{shs1}\bhh^{\prime}_1,~H^{\prime}_2~~\mbox{and}~~
~H^{\prime}_1,~\bhh^{\prime}_2,~~\mbox{where}~~
H^{\prime}_{r}=\frac{\hh^{\prime}_{r}+\alpha_{r}\hh_{r}}
{\sqrt{1+|\alpha_{r}|^2}},~r=1,2. \eeq
The electroweak doublets $H_r$, which remain massless at the GUT
scale, are orthogonal to the $H^{\prime}_{r}$ directions:
\beq H_r=\frac{-\alpha_r^*\hh^{\prime}_r+\hh_r}
{\sqrt{1+|\alpha_r|^2}} \cdot\label{elws}\eeq

\subsection{Yukawa Quasi-Unification Conditions}\label{secWy}

The part of $W$ in \Eref{Wtotal} which includes the Yukawa
interactions of the third family of fermions is given by
\beq W_{\rm Y}=y_{33}F_3\hh
F_3^c+2y'_{33}F_3\hh'F_3^c=y_{33}F_3\lin{\hh_2+2\rho
\hh'_2}{\hh_1+2\rho \hh'_1}F_3^c, \label{Wy}\eeq
where $\rho\equiv y_{33}^{\prime}/y_{33}$ can be made real and
positive by readjusting the phases of $\hh$, $\hh'$, and $H_r$. Also,
note that the factor of two is incorporated in the second term in
the RHS of this equation in order to make $y_{33}^{\prime}$
directly comparable to $y_{33}$, since $\hh_1^{\prime}$ and
$\hh_2^{\prime}$ are proportional to $T^{15}_{\rm c}$, which is
normalized so that the trace of its square equals unity. Solving
Eqs.~(\ref{shs1}) and (\ref{elws}) w.r.t. $\hh_r$ and
$\hh^\prime_r$, we obtain
\beq\label{hh}
\hh_r=\frac{H_r+\alpha^*_rH^{\prime}_r}{\sqrt{1+|\alpha_r|^2}}
~~\mbox{and}~~\hh^\prime_r=\frac{-\alpha_rH_r+H^{\prime}_r}
{\sqrt{1+|\alpha_r|^2}}\cdot~~~\eeq
From Eqs.~(\ref{Wy}) and (\ref{hh}) and using the fact that the
superheavy doublets $H^{\prime}_r$ must have zero VEVs, we can
readily derive the superpotential terms of the MSSM for the third
family fermions as well as the Yukawa interaction of the left
handed third family lepton doublet with $\nu^c_3$:
\bea W_{\rm Y} = -h_t\hu^\tr \openep {Q}_{3} u^c_{3}+h_b\hd^\tr
\openep Q_{3}d^c_{3}+h_\tau\hd^\tr\openep {L}_3e^c_3
-h_{\nu_\tau}\hu^\tr \openep L_3\nu^c_3, \label{wmssm}\eea
where $Q_i=\lin{u_i}{d_i}^\tr$ and
$L_i=\lin{\nu_i}{e_i}^\tr$ are the $SU(2)_{L}$
doublet LH quark and lepton superfields respectively and the
Yukawa coupling constants $h_t$, $h_b$, $h_\tau$, and
$h_{\nu_\tau}$ satisfy a set of generalized asymptotic YQUCs:
\bea \nonumber && h_t(M_{\rm GUT}):h_b(M_{\rm GUT}):h_\tau(M_{\rm
GUT}):h_{\nu_\tau}(M_{\rm GUT})=\\
&& \left|\frac{1-{\rho\alpha_2/\sqrt{3}}}
{\sqrt{1+|\alpha_2|^2}}\right|:
\left|\frac{1-{\rho\alpha_1/\sqrt{3}}}
{\sqrt{1+|\alpha_1|^2}}\right|:
\left|\frac{1+\sqrt{3}\rho\alpha_1}
{\sqrt{1+|\alpha_1|^2}}\right|:
\left|\frac{1+\sqrt{3}\rho\alpha_2}
{\sqrt{1+|\alpha_2|^2}}\right|.~~~~ \label{quasi} \eea
These conditions depend on two complex ($\alpha_1$, $\alpha_2$)
and one real and positive ($\rho$) parameter. For natural values
of $\rho$, $\alpha_1$, and $\alpha_2$, i.e. for values of these
parameters which are of order unity and do not lead to unnaturally
small numerators in the RHS of Eq.~(\ref{quasi}), we expect all
the ratios $h_m/h_n$ with $m,n=t,b,\tau, \nu_\tau$ to be of order
unity. So, exact YU is naturally broken, but not completely lost
since the ratios of the Yukawa coupling constants remain of order
unity, thereby restricting $\tnb$ to rather large values. On the
other hand, these ratios do not have to obey any exact relation
among themselves as in the previously studied
\cite{qcdm,muneg,nova,nova2,yqu,pekino, pekinoa} monoparametric case. As
we show below, this gives us an extra freedom which allows us to
satisfy all the phenomenological and cosmological requirements
with the lightest neutralino contributing to CDM.

\subsection{\boldmath The Peccei-Quinn Symmetry and the
$\mu$ Problem}
\label{secWpq}

The last term $W_{\rm PQ}$ of $W$ in \Eref{Wtotal} is responsible
for the solution of the $\mu$ problem of MSSM. Indeed, an
important shortcoming of MSSM is that there is no understanding of
how the SUSY $\mu$ term with the right magnitude of $|\mu|\sim
10^{2}-10^{3}~{\rm GeV}$ arises. One way \cite{rsym} to solve this
$\mu$ problem is via a PQ symmetry $U(1)_{\rm PQ}$ \cite{pq, pqa,
pqb}, which also solves the strong CP problem. This solution is
based on the observation \cite{kn} that the axion decay constant
$f_{a}$, which is the symmetry breaking scale of $U(1)_{\rm PQ}$,
is (normally) of intermediate value ($\sim 10^{11}-10^{12}~{\rm
GeV}$) and, thus, $|\mu|\sim f_{a}^2 /M_{\rm S}$. The scale
$f_{a}$ is, in turn, of order $(m_{3/2}M_{\rm S})^{1/2}$, where
$m_{3/2}\sim 1~{\rm{TeV}}$ is the gravity-mediated soft SUSY
breaking scale (gravitino mass). In order to implement this
solution of the $\mu$ problem in our model, we introduce
\cite{rsym} a pair of gauge singlet superfields $N$ and $\bar{N}$
(see \Tref{tab1}) with the following nonrenormalizable couplings
in the superpotential -- for an alternative class of
superpotentials, see  \citen{laz4, laz4a, laz4b, laz4c, laz4d}:
\beq W_{\rm PQ} = \lambda_{\mu} \frac{N^2 \hh^2}{M_{\rm S}}
+\lambda^\prime_{\mu} \frac{N^2 \hh^{\prime2}}{M_{\rm S}}+
\lambda_{\rm PQ} \frac{N^2 \bar{N}^2}{M_{\rm S}}\cdot
\label{eq:superpot} \eeq
Here, $\lambda_{\rm PQ}$ is taken real and positive by redefining
the phases of the superfields $N$ and $\bar{N}$. After SUSY breaking,
the $N^2\bar N^2$ term leads to the scalar potential
\begin{eqnarray}
V_{\rm PQ}&=&\left(m_{3/2}^2 +4\lambda_{\rm
PQ}^2\left|\frac{N\bar{N}}{M_{\rm S}}\right|^2\right)
\left[(|N|-|\bar{N}|)^2+2|N||\bar{N}|\right] \nonumber \\ &
&+2|A|m_{3/2}\lambda_{\rm PQ}\frac{|N\bar{N}|^2}{M_{\rm S}}
{\rm{cos}}(\epsilon+2\theta+2\bar{\theta}), \label{eq:pqpot}
\end{eqnarray}
where $A$ is the dimensionless coefficient of the soft SUSY
breaking term corresponding to the superpotential term
$N^2\bar{N}^2$ and $\epsilon$, $\theta$, $\bar{\theta}$ are the
phases of $A$, $N$, $\bar{N}$ respectively. Minimization of
$V_{\rm PQ}$ then requires $|N|=|\bar{N}|$,
$\epsilon+2\theta+2\bar{\theta}=\pi$ and $V_{\rm PQ}$ takes the
form
\begin{equation}
V_{\rm PQ}=2|N|^2m_{3/2}^2\left(4\lambda_{\rm PQ}^2\frac{|N|^4}
{m_{3/2}^2M_{\rm S}^2}-|A|\lambda_{\rm
PQ}\frac{|N|^2}{m_{3/2}M_{\rm S}} +1\right). \label{eq:pqpotmin}
\end{equation}
For $|A|>4$, the absolute minimum of the potential is at
\begin{equation}
|\langle N\rangle|=|\langle\bar{N}\rangle|\equiv
\frac{f_a}{2}=\sqrt{m_{3/2}M_{\rm S}}\
\sqrt{\frac{|A|+\sqrt{|A|^2-12}}{12\lambda_{\rm PQ}}}\sim
\sqrt{m_{3/2}M_{\rm S}}. \label{eq:solution}
\end{equation}
The $\mu$ term is generated predominantly via the terms $N^2\hh^2$
and $N^2\hh^{\prime2}$ in Eq. (\ref{eq:superpot}) with
$|\mu|\sim|\langle N\rangle|^2/M_{\rm S}$, which is of the right
magnitude.

The potential $V_{\rm PQ}$ also has a local minimum at
$N=\bar{N}=0$, which is separated from the global PQ minimum by a
sizable potential barrier, preventing a successful transition from
the trivial to the PQ vacuum. This situation persists at all
cosmic temperatures after reheating, as has been shown \cite{jean}
by considering the one-loop temperature corrections \cite{jackiw,jackiw2}
to the scalar potential. We are, thus, obliged to assume that,
after the termination of inflation, the system emerges with the
appropriate combination of initial conditions so that it is led
\cite{curvaton} to the PQ vacuum.

\subsection{Proton Stability} \label{sec:prot}

One can assign baryon number $B=1/3~[-1/3]$ to all the color
triplets [antitriplets] of the model, which exist not only in
$F, F^c$, but also in $H^c, \bar{H}^c, G$, and the extra Higgs
superfields. Lepton number ($L$)
can then be defined via $B-L$. Before including the extra Higgs
superfields in Table~\ref{tab1}, baryon and lepton number
violation originates from the terms \cite{jean}:
\begin{equation}
F^c F^c H^c H^c,~~ F F \bar{H}^c \bar{H}^c \hh \hh,~~F F \bar{H}^c
\bar{H}^c \bar{N}^2  \label{terms}
\end{equation}
(as well as the terms containing the combinations $(H^c)^4$,
$(\bar{H}^c)^4$), which give couplings like $u^c d^c d^c_H
\nu^c_H$ (or $u^c d^c u^c_H e^c_H$),\ $u d \bar{d}^c_H
\bar{\nu}^c_H$ (or $u d \bar{u}^c_H \bar{e}^c_H$) with appropriate
coefficients. Also, the terms $G H^c H^c$ and $G \bar{H}^c
\bar{H}^c$ give rise to the $B$ (and $L$) violating couplings $g^c
u^c_H d^c_H$,\ $\bar{g}^c \bar{u}^c_H \bar{d}^c_H$. All other
combinations are $B$ (and $L$) conserving since all their
$SU(4)_{\rm c}$ {$\bf 4$}'s are contracted with ${\bf \bar4}$'s.

The dominant contribution to proton decay comes from effective
dimension five operators generated by one-loop diagrams with two
of the $u^c_H$, $d^c_H$ or one of the $u^c_H$, $d^c_H$ and one of
the $\nu^c_H$, $e^c_H$ circulating in the loop. The amplitudes
corresponding to these operators are estimated to be at most of
order $m_{3/2}M_{\rm GUT}/M_{\rm S}^3 \lesssim 10^{-34}~
{\rm{GeV}}^{-1}$. This makes the proton practically stable.

After the inclusion of the superfields $\hh^{\prime}$ and
$\bar{\hh}^{\prime}$, the couplings
\beq FF\bar{H}^c\bar{H}^c\hh\hh^{\prime},
~FF\bar{H}^c\bar{H}^c\hh^{\prime}\hh^{\prime} \label{couplings}
\eeq
(as well as the new couplings containing arbitrary powers of the
combinations $(H^c)^4$, $(\bar{H}^c)^4$) give rise \cite{qcdm} to
additional $B$ and $L$ number violation. However, their
contribution to proton decay is subdominant to the one arising
from the terms of Eq.~(\ref{terms}). One can further show
\cite{qcdm} that the inclusion of the superfields $\phi$,
$\bar\phi$, $\phi'$, and $\bar\phi'$  also gives a subdominant
contribution to the proton decay.

\section{The Resulting CMSSM} \label{masspar}

Below $M_{\rm GUT}$, the particle content of our models reduces to
this of MSSM -- modulo SM singlets. The Yukawa coupling constants
of the models satisfy \Eref{quasi}. To avoid complications with
the seesaw mechanism, we neglect in our analysis the effects from
$h_{\nu_\tau}$ on the \emph{renormalization group} (RG) running
and the SUSY spectrum, although its impact can be sizable
\cite{nucmssm}. We
specify below the adopted SUSY breaking scheme (\Sref{sec:soft}),
describe the derivation of the (s)particle spectrum paying special
attention to the two lightest sparticle mass eigenstates
(\Sref{sec:lsp}), and discuss the fermion masses (\Sref{sec:lsp}).

\subsection{Soft SUSY Breaking in the CMSSM} \label{sec:soft}

The relevant gravity-mediated soft SUSY-breaking terms in the scalar
potential are
\bea V_{\rm soft}&=&
m_{F}^{2}|F|^2-A_th_tH_2^\tr\openep \tilde Q_3 \tilde u_3^c
+A_b h_b H_1^\tr\openep \tilde Q_3\tilde d_3^c
+A_\tau h_\tau H_1^\tr\openep \tilde L_3\tilde e_3^c\nonumber\\
&& -B\mu
H_1^\tr\openep H_2+ {\rm h.c.}~~~\mbox{with}~~~F
=H_1,~H_2,~\tilde L_i,~\tilde e^c_i,~\tilde Q_i,~\tilde u^c_i,
~\tilde d^c_i, \label{vsoft} \eea
where tilde denotes the superpartner and the $A$-terms for
the two light families, although included, are not shown
explicitly. The soft gaugino mass terms in the Lagrangian are
\begin{equation}
\mathcal{L}_{\rm gaug}=\frac{1}{2}\left( M_1\tilde B\tilde
B+M_2\sum_{r=1}^{3}\tilde W_r \tilde W_r+M_3\sum_{a=1}^{8}\tilde
g_a\tilde g_a+{\rm h.c.}\right), \label{gaugino}
\end{equation}
where $\tilde B$, $\tilde W_r$, and $\tilde g_a$ are the bino,
winos, and gluinos respectively.

The SUSY-breaking parameters $A_t,A_b,A_\tau,B$, and $M_\alpha$
($\alpha=1,2,3$) are all of the order of the soft SUSY-breaking scale
$\sim 1~{\rm TeV}$, but are otherwise unrelated in the general
case. However, if we assume that soft SUSY breaking is mediated by
\emph{minimal supergravity} (mSUGRA), i.e. supergravity with minimal
K\"{a}hler potential and minimal gauge kinetic function, we obtain
soft terms which are universal `asymptotically' (i.e. at $M_{\rm GUT}$).
More explicitly, mSUGRA implies

\begin{itemize}

\item a common mass $M_{1/2}$ for gauginos:
\beq M_1(M_{\rm GUT})= M_2(M_{\rm GUT})= M_3(M_{\rm
GUT})=M_{1/2}\label{Muni},~~~~~~~~~\eeq

\item  a common mass $m_0$ for scalars:
\beqs\bea &&\mbox{Sleptons:}~~ m_{\tilde L_i}(M_{\rm GUT})=
m_{\tilde e^c_i}(M_{\rm GUT})=m_0,\\
&&\mbox{Squarks:}~~ m_{\tilde Q_i}(M_{\rm GUT})=m_{\tilde u^c_i}
(M_{\rm GUT})=m_{\tilde d^c_i}(M_{\rm GUT})=m_0,\\
&&\mbox{Higgs:}~~ m_{H_1}(M_{\rm GUT})=m_{H_2}(M_{\rm GUT})=m_0,
\eea\eeqs
\item a common trilinear coupling constant $A_0$:
\beq A_t(M_{\rm GUT})=A_b(M_{\rm GUT})=A_{\tau}(M_{\rm GUT})= A_0,
~~~~~~~~~~~~~~~~
\label{Auni} \eeq
where again only the third family trilinear coulpings are shown
explicitly.
\end{itemize}
The MSSM supplemented by universal boundary conditions is
generally called constrained MSSM (CMSSM) \cite{Cmssm, Cmssma,
Cmssmb, Cmssmc}. It is true that the mSUGRA implies two more
asymptotic relations: $B_0=A_0-m_0$ and $m_0=m_{3/2}$, where
$B_0=B(M_{\rm GUT})$ and $m_{3/2}$ is the (asymptotic) gravitino
mass. These extra conditions are usually not included in the
CMSSM. Imposing them, we get the so-called very CMSSM
\cite{vCmssm, vCmssma}, which is a very restrictive version of
MSSM and will not be considered further here. Therefore, the free
parameters of our model are
\begin{equation}
\sign\mu,~~\tan\beta,~~\Mg,~~m_0,~~\mbox{and}~~A_0, \label{param}
\end{equation}
where $\sign\mu$ is the sign of $\mu$ and
$\tnb=\vev{H_2}/\vev{H_1}$.

In order to proceed with the investigation of the parameter space
of the CMSSM, we integrate the two-loop RG equations
for the gauge and Yukawa coupling constants and the one-loop ones
for the soft SUSY breaking parameters between the SUSY GUT scale
$M_{\rm GUT}$ and a common SUSY threshold
\beq M_{\rm SUSY} \simeq(m_{\tilde t_1}m_{\tilde
t_2})^{1/2}~~\mbox{($\tilde t_{1,2}$ are the stop mass
eigenstates)}\label{msusy}\eeq
determined in consistency with the SUSY spectrum. At $M_{\rm
SUSY}$, we impose radiative EWSB and express the values of the
parameters $\mu$ (up to its sign) and $B$ (or, equivalently, the
mass $m_A$ of the CP-odd neutral Higgs boson $A$) at $M_{\rm
SUSY}$ in terms of the other input parameters by minimizing the
tree-level RG improved potential \cite{pierce, piercea} at $M_{\rm
SUSY}$. The resulting conditions are
\begin{equation}
\mu^2=\frac{m^2_{H_1}-m^2_{H_2}\tan^2{\beta}}
{\tan^2{\beta}-1}-\frac{1}{2} M^2_Z , \quad \sin 2\beta=\frac{2
B\mu}{m_{1}^2+m_{2}^2}\equiv \frac{2B\mu}{m_A^2}, \label{mu}
\end{equation}
where $m_A^2=m_{1}^2+m_{2}^2$ with $m_{1}^2=m_{H_{1}}^2+\mu^2$ and
$m_{2}^2=m_{H_{2}}^2+\mu^2$. We could improve the accuracy of
these conditions by including the full one-loop radiative
corrections to the potential from Ref.~\citen{pierce, piercea} at
$M_{\rm SUSY}$. It is shown \cite{thresholdbcdpt,
thresholdbcdpta}, however, that the corrections to $\mu$ and $m_A$
from the full one-loop effective potential are minimized by our
choice of $M_{\rm SUSY}$. So, we will not include these
corrections, but rather use this variable SUSY threshold which
gives a much better accuracy than a fixed one.

We then evaluate the SUSY spectrum by employing the publicly
available calculator {\tt SOFTSUSY} \cite{Softsusy} and
incorporate the SUSY corrections to the $b$ and $\tau$ mass
\cite{pierce, piercea}. The corrections to the $b$-quark mass
arise from sbottom-gluino (mainly) and stop-chargino loops
\cite{copwa, copwb,pierce, piercea} and have the sign of $\mu$ --
with the standard sign convention of Ref.~\citen{sugra}. Less
important but not negligible (almost 4$\%$) are the SUSY
corrections to the $\tau$-lepton mass originating \cite{pierce,
piercea} from sneutrino-chargino (mainly) and stau-neutralino
loops and leading \cite{qcdm, muneg} to a small decrease of
$\tan\beta$. From $M_{\rm SUSY}$ to $M_Z$, the running of the
gauge and Yukawa coupling constants is continued using the SM RG
equations.

\subsection{The LSP and the Next-to-LSP} \label{sec:lsp}

We now focus on the two lightest sparticles whose mass proximity
plays a crucial role in constructing a viable CDM scenario --
see \Sref{sec:cosmo}. In particular, the role of the LSP can be
played by the lightest neutralino $\tilde{\chi}$, whereas the
\emph{next-to-LSP} (NLSP) can be the lightest stau mass eigenstate
$\tilde\tau_2$. Moreover, for presentation purposes, $M_{1/2}$ and
$m_0$ can sometimes be replaced \cite{cdm,cdm2} by the LSP mass $\mx$
and the relative mass splitting $\Delta_{\tilde\tau_2}$ between
$\xx$ and $\staub$ defined as follows:
\beq \Delta_{\tilde\tau_2}=(m_{\tilde\tau_2}-m_{\rm LSP})/m_{\rm
LSP}.\label{Dst}\eeq

\par
The LSP mass $\mx$ can be obtained by diagonalizing the mass
matrix of the four neutralinos, which is
\begin{eqnarray}
{\llgm\bem M_1 & 0 & -M_Z s_W\cos\beta & M_Z s_W\sin\beta
 \cr 0 & M_2 & M_Z c_W\cos\beta & -M_Z c_W\sin\beta
 \cr -M_Z s_W\cos\beta & M_Z c_W\cos\beta & 0 & -\mu
 \cr M_Z s_W\sin\beta & -M_Z c_W\sin\beta & -\mu & 0
\cr\eem\rrgm} \label{neutralino}
\end{eqnarray}
in the $(-i\tilde B, -i\tilde W_3, \tilde H_1, \tilde H_2)$ basis.
Here, $s_W=\sin \theta_W$, $c_W=\cos \theta_W $, and $M_1$, $M_2$
are the masses of $\tilde B$, $\tilde W_3$ in Eq.~(\ref{gaugino}).
In the CMSSM and for most of the parameter space, $\mx\simeq M_1$
and, thus, $\xx$ turns out to be an almost pure bino $\tilde B$.

The evolution of the gaugino masses $M_\alpha$ at one loop can be
easily found  by solving the relevant RG equations \cite{Cmssm,
Cmssma, Cmssmb, Cmssmc}, which admit an exact solution:
\beq M_\alpha(Q)=M_{1/2}\frac{g_\alpha(Q)}{g_{\rm GUT}}=
M_{1/2}\left(1-\frac{b_\alpha}{8\pi^2}\ln\frac{Q}{M_{\rm GUT}}
\right)^{-1},\label{M12}\eeq
where $(b_\alpha)=(33/5,1,-3)$, $g_\alpha$ are the gauge coupling
constants associated with the gauge groups $U(1)_{\rm Y}$,
$SU(2)_{L}$, and $SU(3)_{\rm c}$ respectively, and $g_{\rm GUT}
\simeq1/24$ is the common value of the $g_\alpha$'s at the
GUT scale $M_{\rm GUT}\simeq2\times10^{16}~\GeV$. After a direct
computation, we obtain $\mx(M_{\rm SUSY})\simeq0.45 M_{1/2}$.

The lightest stau mass eigenstate
$\tilde\tau_2$ can be obtained by diagonalizing the stau
mass-squared matrix
\begin{eqnarray}
{\llgm\bem m_{\tau}^2+m_{L}^2+M_Z^2 (-\frac{1}{2}+s_W^2)\cos
2\beta & m_{\tau}(A_{\tau}-\mu\tan\beta) \cr
m_{\tau}(A_{\tau}-\mu\tan\beta) & m_{\tau}^2+m_{E}^2-M_Z^2
s_W^2\cos 2\beta\cr\eem\rrgm}\label{stau}
\end{eqnarray}
in the gauge basis ($\tilde\tau_{\rm L},\ \tilde\tau_{\rm R}$).
Here, $m_{\tilde\tau_{\rm L[R]}}$ is the soft SUSY-breaking mass
of the left [right] handed stau $\tilde\tau_{\rm L[R]}$,
$m_{\tau}$ the tau-lepton mass, and the simplifying notation
$m_{L}\equiv m_{\tilde L_3}$ and $m_{E}\equiv m_{\tilde e^c_3}$
for the third generation soft SUSY-breaking slepton masses is
used. The stau mass eigenstates are
\begin{eqnarray}
{\llgm\bem\tilde\tau_1 \cr \tilde\tau_2 \cr\eem\rrgm} & = &
{\llgm\bem\cos\theta_{\tilde{\tau}} & \sin\theta_{\tilde{\tau}}
\cr -\sin\theta_{\tilde{\tau}} & \cos\theta_{\tilde{\tau}}
\cr\eem\rrgm} {\llgm\bem\tilde\tau_{\rm L} \vspace{0.3cm}\cr
\tilde\tau_{\rm R} \cr\eem\rrgm}, \label{eigen}
\end{eqnarray}
where $\theta_{\tilde{\tau}}$ is the $\tilde\tau_{\rm
L}-\tilde\tau_{\rm R}$ mixing angle. The large values of the $b$ and
$\tau$ Yukawa coupling constants, implied by the YQUCs, cause soft
SUSY-breaking masses of the third generation squarks and sleptons
to run (at low energies) to lower physical values than the
corresponding masses of the first and second generation.
Furthermore, the large values of $\tan\beta$, implied again by YQUCs,
lead to large off-diagonal mixings in the sbottom and stau
mass-squared matrices. These effects reduce further the physical
$\mst$, which becomes easily the NLSP.

\begin{figure}[!t]\vspace*{-.35in}
\centerline{\epsfig{file=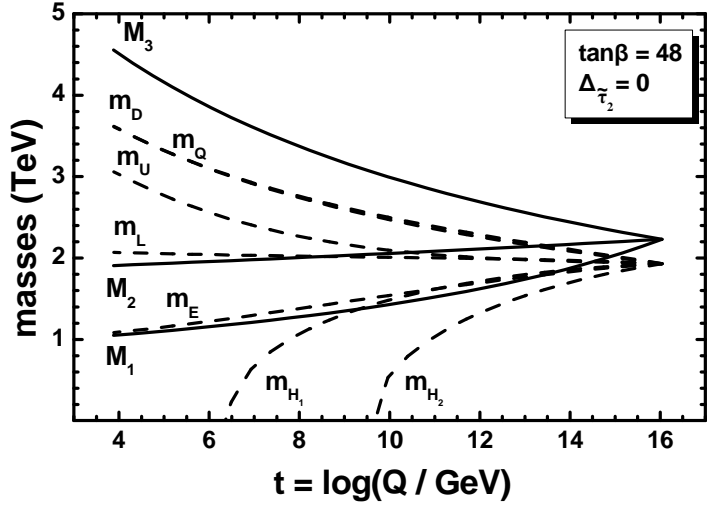,angle=-90,width=11cm}}
\hfill \caption{\sl\ftn The RG evolution from $Q=M_{\rm SUSY}$ to
$Q=M_{\rm GUT}$ of the soft SUSY-breaking masses of the Higgs bosons
($m_{H_1}$ and $m_{H_2}$ ), the third generation scalars
($m_U, m_D, m_E, m_Q$, and $m_L$), and the gauginos
($M_1, M_2$, and $M_3$) for $\tnb=48,~\Dst\simeq0,~\AMg=-1.4$, and
$\Mg=2.2~\TeV$.}\label{masses}
\end{figure}

In \Fref{masses}, an example of the RG running from $Q=M_{\rm
SUSY}$ to $Q=M_{\rm GUT}$ of the soft SUSY-breaking masses of the
Higgs and the third generation scalars as well as the gauginos is
shown. Here, we extend the simplifying notation for the soft
masses to include the masses of the third generation squarks too:
$m_Q\equiv m_{\tilde Q_3}$, $m_U\equiv m_{\tilde u^c_3}$, and
$m_D\equiv m_{\tilde d^c_3}$ and take
$\tnb=48,~\Dst\simeq0,~\AMg=-1.4$, and $\Mg=2.2~\TeV$ resulting to
$\mu=2.78~\TeV$. It is rather amazing that, from so few inputs,
all of the masses of the SUSY particles can be determined. One
characteristic feature of the spectrum which is obvious from this
figure is that the colored sparticles are typically the heaviest
particles. This is due to the large enhancement of their masses
originating from the $SU(3)_c$ gauge coupling $g_3$ in the RG
equations. Also, one sees that $\xx$ can typically play the role
of the LSP. Most importantly, though, one notices that $m_{H_1}^2$
and $m_{H_2}^2$ reach zero and then become negative triggering the
EWSB. In particular, we obtain $m_1^2(M_{\rm
SUSY})=4.41~\TeV^2>0$, but $m_2^2(M_{\rm SUSY})=-2.2~\TeV^2<0$ and
so the quadratic part of the scalar potential for the electrically
neutral components of the Higgs fields becomes indefinite -- see
e.g. \cref{book} -- causing the EWSB.

\subsection{The Masses of the Fermions} \label{frm}

The masses of the fermions of the third generation play a crucial
role in the determination of the evolution of the Yukawa coupling
constants. For the $b$-quark mass, we adopt as an input parameter
in our analysis the $\overline{\rm MS}$ $b$-quark mass, which at
$1-\sigma$ is \cite{pdata}
\beq m_b \lf m_b\rg^{\overline{\rm MS}}=4.19^{+0.18}_{-0.06}~\GeV.
\eeq
This range is evolved up to $M_Z$ using the central value
$\alpha_s(M_Z)=0.1184$ \cite{pdata} of the strong fine structure
constant at $M_Z$ and then converted to the ${\rm \overline{DR}}$
scheme in accordance with the analysis of Ref.~\citen{baermb,
baermba}. We obtain, at $95\%$ c.l.,
\beq 2.745\lesssim  m_b(M_Z)/{\rm GeV}\lesssim 3.13
\label{mbrg}\eeq
with the central value being $m_b(M_Z)=2.84~\GeV$. For the
top-quark mass, we use the central pole mass ($M_t$) as an input
parameter \cite{mtmt, mtmta}:
\beq M_t=173~\GeV~~\Rightarrow~~m_t(m_t)=164.6~\GeV\eeq
with $m_t(m_t)$ being the running mass of the top quark. We also
take the central value $m_{\tau}(M_Z) = 1.748~\GeV$ \cite{baermb,
baermba} of the ${\overline{\rm DR}}$ tau-lepton mass at $M_Z$.

\begin{figure}[!t]
\vspace*{-.35in}
\centerline{\epsfig{file=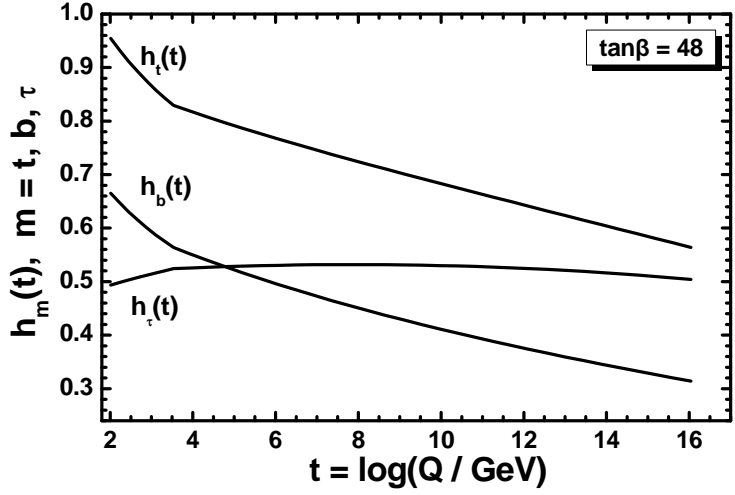,angle=-90,width=11cm}}
\hfill \caption{\sl\ftn The RG evolution from $Q=M_Z$ to $Q=M_{\rm
GUT}$ of the third generation Yukawa coupling constants for
$\tnb=48,~\AMg=-1.4$, $\Mg=2.27~\TeV$, and $m_0=1.92~\TeV$.}
\label{yuk}
\end{figure}

In Fig.~\ref{yuk}, we present an example of a third generation
Yukawa coupling constant RG running from $M_{\rm GUT}$ to $M_Z$
for $\tnb=48,~\AMg=-1.4$, $\Mg=2.27~\TeV$, and $m_0=1.92~\TeV$.
At $M_{\rm GUT}$, we have $h_t/h_\tau=1.117$, $h_b/h_\tau=0.623$, and
$h_t/h_b=1.792$. This is, actually, the first out of the four cases
of Table~\ref{tab:spectrum} (see below).  As we show in \Sref{viol},
these ratios can be naturally obtained from the YQUCs in \Eref{quasi}.
The kinks on the various curves correspond to the point where the
MSSM RG equations are replaced by the SM ones. We observe that
$h_\tau$ is greater than $h_b$ but lower than $h_t$ at $\Mgut$.

\section{Phenomenological Constraints}
\label{sec:pheno}

The model parameters are restricted by a number of
phenomenological and cosmological constraints, which are evaluated
by employing the latest version of the publicly available code
{\tt micrOMEGAs} \cite{micro, microa}. We briefly discuss below
the phenomenological constraints paying special attention to those
which are most relevant to our investigation.

\subsection{The Higgs Boson Mass}\label{phenoh}

According to recent independent announcements from the ATLAS
\cite{atlas} and the CMS \cite{cms} experimental teams at the LHC
-- see also \cref{cdf} -- a discovered particle, whose behavior
is consistent with the SM Higgs boson, has a mass around
$125-126~\GeV$. More precisely, the reported mass is
\beq m_h=\left\{
\begin{array}{rl}126.0 \pm 0.4 \>\>\mbox{(stat)}\pm \>\> 0.4\>\>\mbox{(sys)} \>\>\GeV &
\mbox{ATLAS},\\
125.3 \pm 0.4\>\> \mbox{(stat)}\pm \>\> 0.5\>\> \mbox{(sys)}
\>\>\GeV & \mbox{CMS}.
\end{array}
\right.\eeq
In the absence of a combined analysis of the ATLAS and CMS
data and allowing for a theoretical uncertainty of
$\pm1.5~\GeV$, we construct a $2-\sigma$ range for $m_h$ adding in
quadrature the various experimental and theoretical uncertainties
and taking the upper [lower] bound from the ATLAS [CMS] results:
\beq 122\lesssim m_h/\GeV\lesssim129.2.\label{mhb} \eeq
This restriction is applied to the mass $m_h$ of the light CP-even
Higgs boson $h$ of MSSM. For the calculation of $m_h$, we use the
package {\tt SOFTSUSY} \cite{Softsusy}, which includes the full
one-loop SUSY corrections and some zero-momentum two-loop
corrections \cite{2loops, 2loopsa, 2loopsb, 2loopsd}. The results
are well tested \cite{comparisons, comparisonsa} against other
spectrum calculators.

\begin{figure}[!t]
\vspace*{-.35in}
\centerline{\epsfig{file=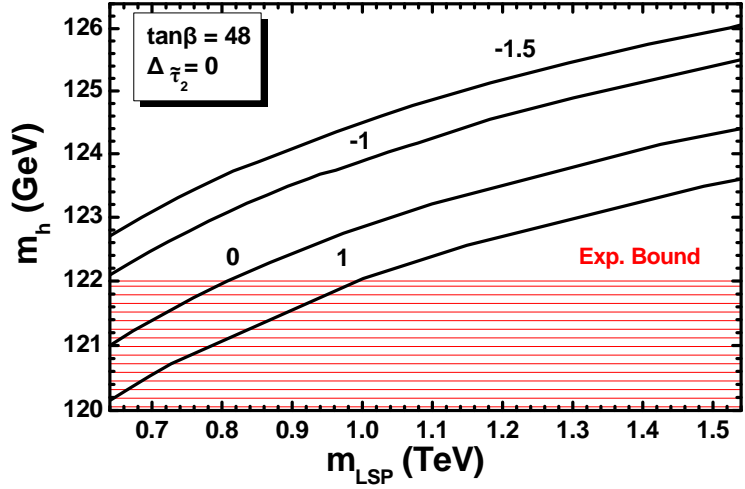,angle=-90,width=11cm}}
\hfill \caption{\sl\ftn The mass $m_h$ of the MSSM Higgs boson as
a function of $m_{\rm LSP}$ for $\tnb=48,~\Dst\simeq0$, and
various $\AMg$'s indicated in the plot. The (red) region is
excluded by the lower bound in Eq.~(\ref{mhb}).}\label{mhf}
\end{figure}

In \Fref{mhf}, we depict $m_h$ as a function of $m_{\rm LSP}$ for
$\tnb=48,~\Dst\simeq0$, and $~\AMg=1,~0,~-1$, and $-1.5$. We notice
that $m_h$ increases with $\mx$ and as $\AMg$ decreases to values
lower than zero. This occurs, since the off-diagonal elements of the
mass-squared matrix of the stop quarks, which contribute to the corrections
to $m_h$, are maximized for $\AMg<0$. As a consequence, the bound on
$\mx$ for $\AMg<0$ turns out to be less restrictive.

\subsection{\boldmath The Branching Ratio $\bmm$}\label{phenoa}

The rare decay $B_s\to \mu^+\mu^-$ occurs via $Z$ penguin and box
diagrams in the SM and, thus, its branching ratio is highly
suppressed. The SUSY contribution, though, originating \cite{bsmm,
bsmma, bsmmb, bsmmc, bsmmd, mahmoudi} from neutral Higgs bosons in
chargino-, $H^\pm$-, and $W^\pm$-mediated penguins behaves as
$\tan^6\beta/m^4_A$ and hence is particularly important for large
$\tan\beta$'s. We impose here the following $95\%$ c.l. bound
\cite{lhcb, lhcba}
\beq \bmm\lesssim4.2\times10^{-9} \label{bmmb}, \eeq
which is significantly reduced relative to the previous
experimental upper bound \cite{bmmexpold}. This bound implies a
lower bound on $\mx$, since $\bmm$ decreases as $\mx$ increases.
Note that, very recently, the LHCb collaboration reported
\cite{LHCb12} a first evidence for the decay $B_s\to\mu^+\mu^-$
yielding the following two sided $95\%$ c.l. bound
\beq 1.1\lesssim\bmm/10^{-9}\lesssim6.4. \label{bmmb12} \eeq
In spite of this newer experimental upper bound on $\bmm$, we
adopt here the much tighter upper bound on $\bmm$ in \Eref{bmmb},
since it is a combined result \cite{bsmumucombined} of the ATLAS,
CMS, and LHCb experiments and, thus, more realistic. As we show
below, the upper bound on the LSP mass $\mx$ which can be inferred
from the lower bound on $\bmm$ in \Eref{bmmb12} does not constrain
the parameters of our model.

\begin{figure}[!t]
\vspace*{-.35in}
\centerline{\epsfig{file=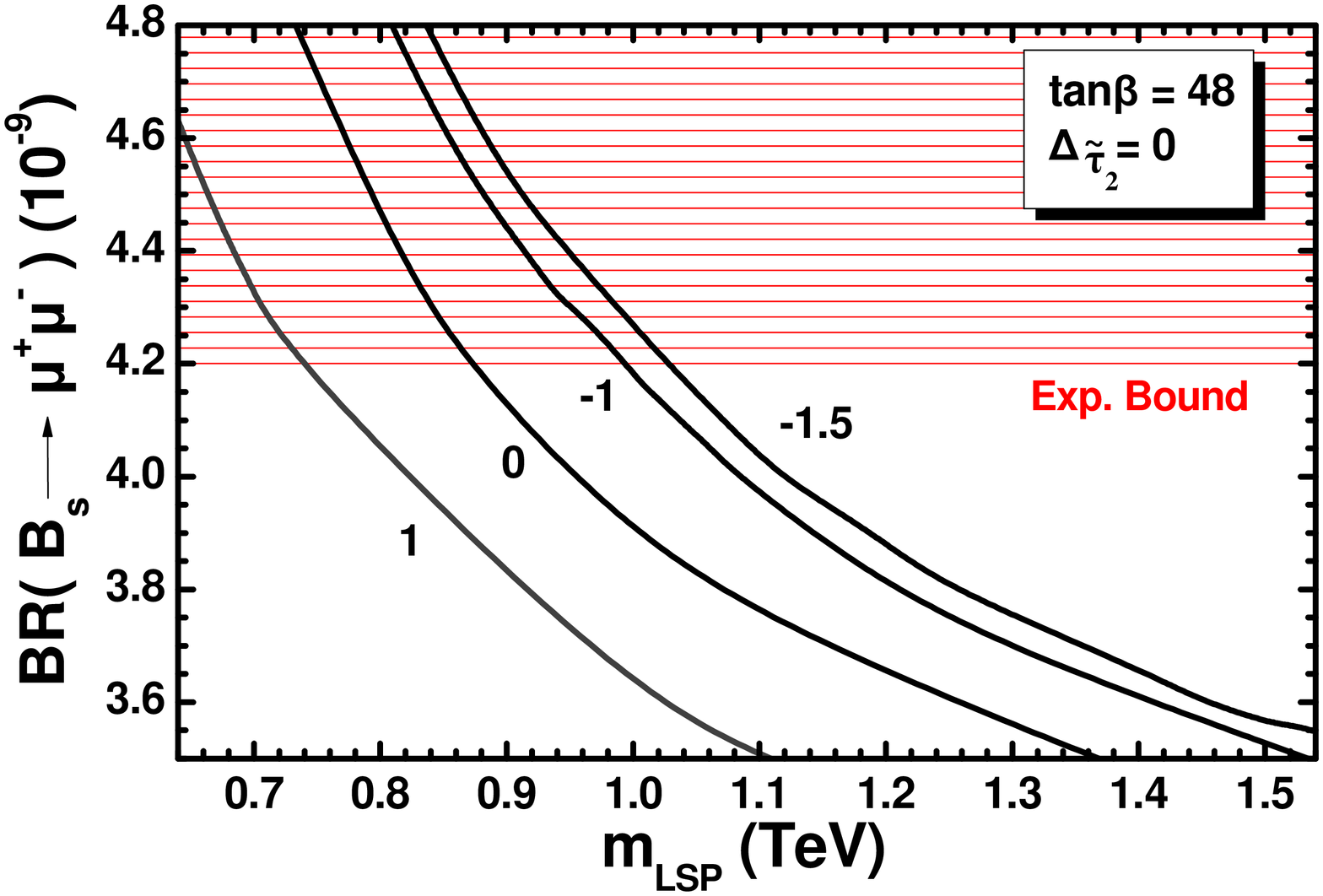,angle=-90,width=11cm}}
\hfill \caption{\sl\ftn $\bmm$ as a function of $m_{\rm LSP}$ for
$\tnb=48,~\Dst\simeq0$, and various $\AMg$'s indicated in the
plot. The experimentally excluded (red) region is also shown.}
\label{bmmf}
\end{figure}

In \Fref{bmmf}, we depict $\bmm$ as a function of $m_{\rm LSP}$ for
$\tnb=48,~\Dst\simeq0$ and $\AMg=1,~0,~-1$, and $-1.5$. We observe
that $\bmm$ decreases as $\mx$ and $\AMg$ increase. Therefore,
for $\AMg<0$, which is favored by the data on $m_h$, the bound on
$\mx$ from \Eref{bmmb} is more restrictive than for $\AMg>0$.

\subsection{\boldmath The Branching Ratio $\bsg$}
\label{phenob}

Combining in quadrature the experimental and theoretical errors in
the most recent experimental world average \cite{bsgexp} and the
SM prediction \cite{bsgSM} for the branching ratio $\bsg$ of the
process $b\to s\gamma$, we obtain the following constraints
at $95\%$ c.l.:
\beq 2.84\times 10^{-4}\lesssim \bsg \lesssim 4.2\times 10^{-4}.
\label{bsgb} \eeq
The computation of $\bsg$ in the {\tt micrOMEGAs} package
presented in \cref{microbsg} includes \cite{nlobsg, nlobsga,
nlobsgb, nlobsgc} \emph{next-to-leading order} (NLO) QCD
corrections to the charged Higgs boson ($H^\pm$) contribution, the
$\tan\beta$ enhanced contributions, and resummed NLO SUSY QCD
corrections. The $H^\pm$ contribution interferes constructively
with the SM contribution, whereas the SUSY contribution interferes
destructively with the other two contributions for $\mu>0$. The SM
contribution plus the $H^\pm$ and SUSY contributions initially
increases with $\mx$ and yields a lower bound on $\mx$ from the
lower bound in Eq.~(\ref{bsgb}). For higher values of $\mx$, it
starts mildly decreasing.

\subsection{\boldmath The Ratio
${\rm R}\lf B_u\to \tau\nu\rg$}\label{phenoc}

The purely leptonic decay $B_u\to \tau\nu$ proceeds via $W^\pm$-
and $H^\pm$-mediated annihilation processes. The SUSY
contribution, contrary to the SM one, is not helicity suppressed
and depends on the mass $m_{H^\pm}$ of the charged Higgs boson
since it behaves \cite{Btn, Btna, mahmoudi} like
$\tan^4\beta/m^4_{H^\pm}$. The ratio $\btn$ of the CMSSM to the SM
branching ratio of the process $B_u\to \tau\nu$ increases with
$\mx$ and approaches unity. It is to be consistent with the
following $95\%$ c.l. range \cite{bsgexp}:
\beq 0.52\lesssim\btn\lesssim2.04\ .\label{btnb} \eeq
A lower bound on $\mx$ can be derived from the lower bound in this
inequality.

\subsection{Muon Anomalous Magnetic Moment}
\label{phenod}

The discrepancy $\delta a_\mu$ between the measured value $a_\mu$
of the muon anomalous magnetic moment and its predicted value in
the SM can be attributed to SUSY contributions arising from
chargino-sneutrino and neutralino-smuon loops. The relevant
calculation is based on the formulas of Ref.~\citen{gmuon}. The
absolute value of the result decreases as $\mx$ increases and its
sign is positive for $\mu>0$. On the other hand, the calculation
of $a^{\rm SM}_\mu$ is not yet stabilized mainly because of the
ambiguities in the calculation of the hadronic vacuum-polarization
contribution. According to the evaluation of this contribution in
Ref.~\citen{g2davier}, there is still a discrepancy between the
findings based on the $e^+e^-$-annihilation data and the ones
based on the $\tau$-decay data -- however, in \cref{Jen}, it is
claimed that this discrepancy can be alleviated. Taking into
account the more reliable calculation based on the $e^+e^-$ data
\cite{Hagiwara}, the recent complete tenth-order QED contribution
\cite{kinoshita}, and the experimental measurements \cite{g2exp}
of $a_\mu$, we end up with a $2.9-\sigma$ discrepancy
\beq~\delta a_\mu=\lf24.9\pm8.7\rg\times10^{-10}, \label{g2b1}\eeq
resulting to the following $95\%$ c.l. range:
\beq~7.5\times 10^{-10}\lesssim \delta a_\mu\lesssim 42.3\times
10^{-10}. \label{g2b}\eeq
A lower [upper] bound on $\mx$ can be derived from the upper
[lower] bound in \Eref{g2b}. As it turns out, only the upper bound
on $\mx$ is relevant here. Taking into account the aforementioned
computational instabilities and the fact that a discrepancy at the
level of about $3-\sigma$ cannot firmly establish a real deviation
from the SM value, we do not consider this bound as a strict
constraint, but rather restrict ourselves to just mentioning at
which level \Eref{g2b1} is satisfied in the parameter space of the
model allowed by all the other constraints -- cf. \cref{CmssmLhc,
CmssmLhca, CmssmLhcb}.

\section{Cold Dark Matter Considerations} \label{sec:cosmo}

The Lagrangian of MSSM is invariant under a discrete
$\mathbb{Z}^{\rm mp}_2$ `matter parity' symmetry, under which all
`matter' (i.e. quark and lepton) superfields change sign -- see
Table~\ref{tab1}. Combining this symmetry with the $\mathbb{Z}_2$
fermion number symmetry, under which all fermions change sign, we
obtain the discrete $\mathbb{Z}_2$ R-parity symmetry, under which
all SM particles are even, while all sparticles are odd. By virtue
of R-parity conservation, the LSP is stable and, thus, can
contribute to the CDM in the universe. It is important to note that
matter parity is vital for MSSM to avoid baryon- and
lepton-number-violating renormalizable couplings in the
superpotential, which would lead to highly undesirable phenomena
such as very fast proton decay. So, the possibility of having the
LSP as CDM candidate is not put in by hand, but arises naturally
from the very structure of MSSM.

The $95\%$ c.l. range for the CDM abundance, according to the
results of WMAP \cite{wmap}, is
\beq \Omega_{\rm CDM}h^2=0.1126\pm0.0072. \label{cdmba}\eeq
The LSP ($\xx$) can be a viable CDM candidate if its relic abundance
$\Omx$ does not exceed the $95\%$ c.l. upper bound derived from
Eq.~(\ref{cdmba}), i.e.
\beq \Omx\lesssim0.12.\label{cdmb}\eeq
Note that, in accordance with the recently reported \cite{planck}
results from the Planck satellite, the CDM abundance is slightly
larger. This leads to an upper bound on $\Omx$ which is somewhat
less restrictive than the one we use in our calculation. The lower
bound on $\Omx$ is not taken into account in our analysis, since
other production mechanisms \cite{scn, scna, scnb, scnc} of LSPs
may be present too and/or other particles \cite{axino, axinoa,
Baerax, Baeraxa, Baeraxb} may also contribute to the CDM. We
calculate $\Omx$ using the \mcr\ code, which includes accurately
thermally averaged exact tree-level cross sections of all the
(co)annihilation processes \cite{cdm, cdm2, cmssm1, cmssm1a}, treats
poles \cite{cmssm2, cmssm2b, qcdm, nra, nraa} properly, and
uses one-loop QCD and SUSY QCD corrected \cite{copwa, copwb, qcdm,
microbsg} Higgs decay widths and couplings to fermions.

The bound in Eq.~(\ref{cdmb}) strongly restricts the parameters of
the CMSSM, since $\Omx$ generally increases with the mass $\mx$ of
the LSP and so an upper bound on $\mx$ can be derived from this
equation. Actually, in most of the parameter space of the CMSSM,
$\Omega_{\rm LSP}h^2$ turns out to be greater than the bound in
\Eref{cdmb} and can become compatible with this equation mainly in
the following clearly distinguished regions (or `islands') in the
CMSSM parameter space:

\begin{itemize}

\item in the bulk region which appears at low values of $m_0$ and
$M_{1/2}$, where $\xx\xx$ annihilation occurs predominantly via
$t$-channel slepton exchange. This region is now excluded by the
bound in \Eref{mhb}.

\item in the \emph{hyperbolic branch/focus point} (HB/FP) region,
which lies at large values of $m_0$ ($>5~\TeV$), where $|\mu|$
becomes small and the neutralino $\xx$ develops a significant
higgsino component \cite{focus, focusa, focusb, focusc, focusd,
focuse, focusf} (for some details, see Sec.~\ref{hbfp}).

\item in the stau coannihilation tail at low $m_0$'s but almost
any value of $M_{1/2}$, where there is a proximity between the
masses of the LSP and the NLSP, which turns out to be the
$\tilde\tau_2$ for $\tan\beta>10$ \cite{cmssm1, cmssm1a} and not
too large values of $|A_0|$ \cite{drees}. Large $|A_0|$ can
generate a stop coannihilation region. For fixed $m_{\rm LSP}$,
$\Omega_{\rm LSP}h^2$ decreases with $\Delta_{\tilde\tau_2}$,
since the $\tilde\chi\tilde\tau_2$ coannihilations become more
efficient. So the CDM criterion can be used for restricting
$\Delta_{\tilde\tau_2}$ -- see Refs.~\citen{cdm,cdm2,qcdm,nova,nova2}.

\item in the $A$-pole enhanced $\xx\xx$ annihilation funnel for
$\tan\beta>40~[\tan\beta\simeq 30-35]$ for $\mu>0~[\mu<0]$,
where one encounters the presence of a resonance with
\beq \Da\equiv\lf m_A-2m_{\rm LSP}\rg/2\mx\simeq0 \label{Da}\eeq
in the $\xx\xx$ annihilation to down type fermions via a $s$-channel
exchange of an $A$-boson.

\end{itemize}

\begin{figure}[!tb]
\centering
\begin{minipage}{\textwidth}
\epsfig{file=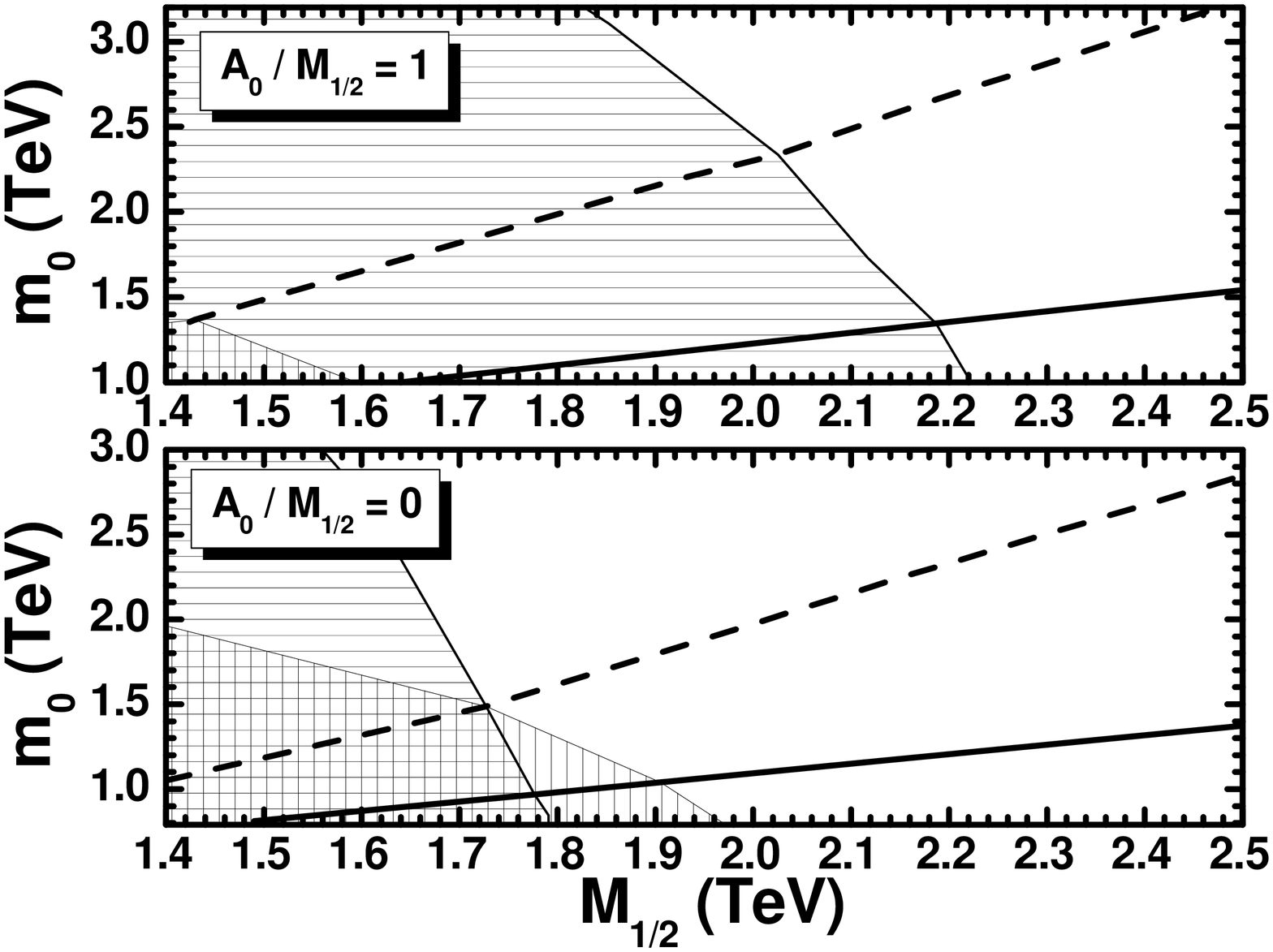,height=2.45in,angle=-90}
\epsfig{file=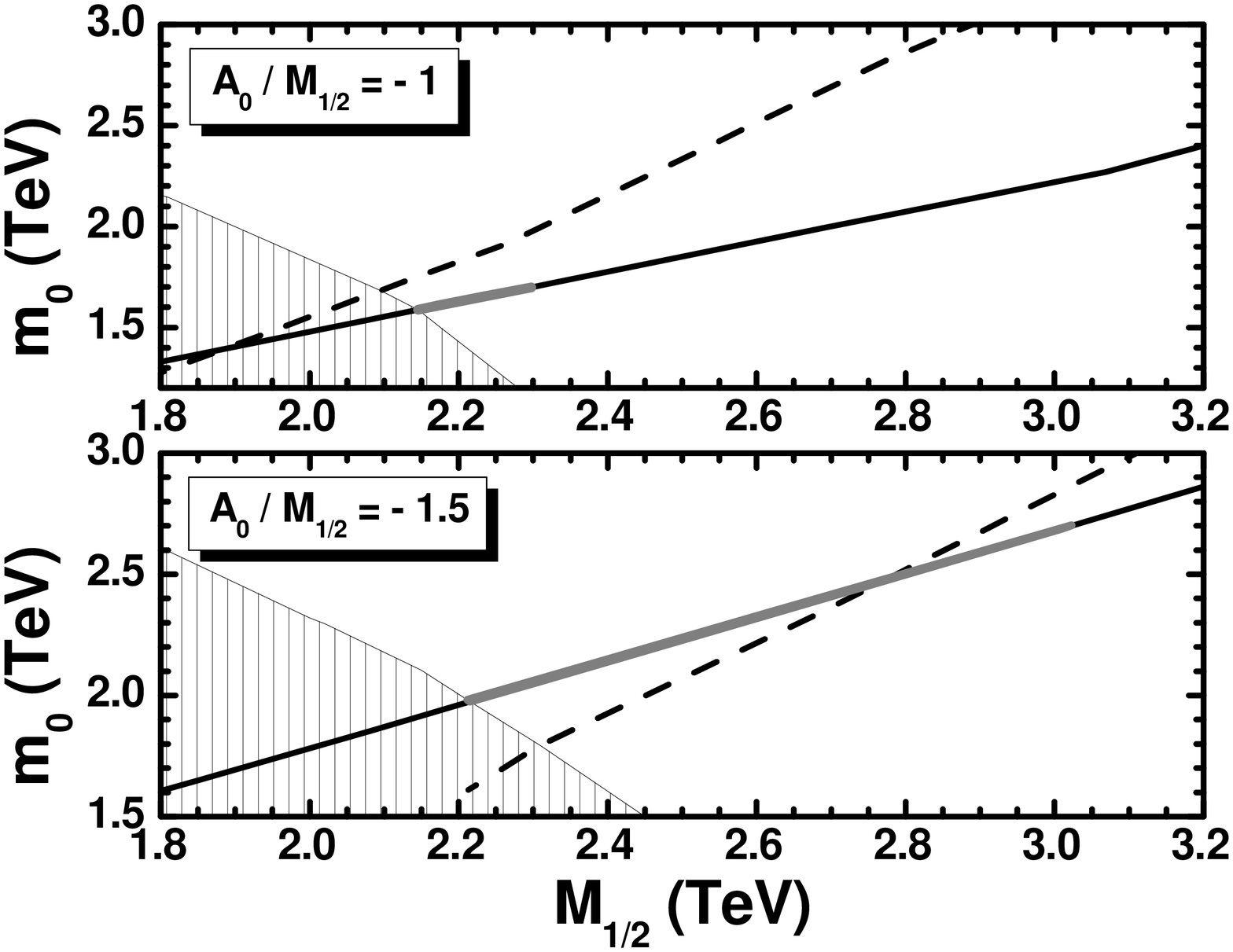,height=2.45in,angle=-90} 
\end{minipage}
\caption{\sl\ftn Relative position of the $\Dst=0$ (solid) line
and the $\Delta_H=0$ (dashed) line in
the $\Mg-m_0$ plane for $\tnb=48$ and various values of $A_0/\Mg$
indicated in the graphs. The vertically [horizontally] hatched regions
are excluded by the bound in \Eref{bmmb} [lower bound in
\Eref{mhb}]. The gray areas are the overall allowed areas.}
\label{fig:DAstau}
\end{figure}

In the region of the CMSSM parameter space which is favored by the
bound in \Eref{mhb} with $\AMg<0$, it is recently recognized
\cite{CmssmLhc, CmssmLhca, CmssmLhcb, gyqu} that there is an area
where two $\Omega_{\rm LSP}h^2$ reduction mechanisms analogous to
the two latter ones mentioned just above cooperate to reduce the
LSP relic abundance below $0.12$. In particular, the lines
$\Dst=0$ and $\Da=0$ can intersect each other in this area,
leading to a resonant enhancement of the $\tilde\chi\tilde\tau_2$
coannihilations. Note that, since $m_A\simeq m_H$, where $m_H$ is
the mass of the heavy CP-even neutral Higgs boson $H$,
$\Da\simeq0$ implies the presence of a resonance $2\mx\simeq m_H$
too. Under these circumstances, the $\tilde\tau_2\tilde\tau_2^*$
coannihilations to $b\bar{b}$ and $\tau\bar\tau$ are enhanced by
the $s$-channel exchange of a $H$-pole -- for the relevant
channels, see, for example, Ref.~\citen{cdm,cdm2}.

In order to pinpoint more precisely this effect, we track in
Fig.~\ref{fig:DAstau} the relative position of the lines $\Dst=0$
and $\Dhp\equiv (m_H-2m_{\rm LSP})/2m_{\rm LSP}=0$ in the
$\Mg-m_0$ plane for $\tan\beta=48$ and various values of
$A_0/M_{1/2}$. The solid [dashed] lines correspond to $\Dst=0$
[$\Dhp=0$]. Also, the vertically [horizontally] hatched regions
are excluded by the bound on $\bmm$ in \Eref{bmmb} [lower bound on
$m_h$ in \Eref{mhb}]. We observe that, for $\AMg = 1$, the lower
bound on $\Mg$ which originates from the lower bound on $m_h$ in
\Eref{mhb} overshadows the one from \Eref{bmmb}. In all other
cases, however, we have the opposite situation. This is consistent
with the fact that, for almost fixed $\Mg$ and $m_0$, $m_h$
increases as $\AMg$ decreases -- see \Fref{mhf}.

From Fig.~\ref{fig:DAstau}, we see that, for $A_0/M_{1/2}=1$ and
0, the $\Dhp=0$ line is far from the part of the $\Dst=0$ line
which is allowed by all the other constraints except the CDM
bound. Consequently, in the neighborhood of this part, the effect
of the $H$-pole is not strong enough to reduce $\Omx$ below 0.12
via $\tilde\tau_2\tilde\tau_2^*$ coannihilations and no overall
allowed area exists. On the contrary, for $A_0/M_{1/2}=-1$, the
$\Dhp=0$ line gets near the otherwise allowed (i.e. allowed by all
the other requirements in \Sref{sec:pheno} without considering the
CDM bound) part of the $\Dst=0$ line and starts affecting the
neighborhood of its leftmost segment, where $\Omx$ becomes smaller
than 0.12 and, thus, an overall allowed (gray) area appears. For
$A_0/M_{1/2}=-1.5$,  the $\Dhp=0$ line moves downwards and
intersects the $\Dst=0$ line. This enhances $H$-pole
$\tilde\tau_2\tilde\tau_2^*$ coannihilation in the neighborhood of
a bigger segment of the otherwise allowed part of the $\Dst=0$
line, where $\Omx$ is reduced below 0.12, and, thus, a bigger
overall allowed (gray) area is generated. For even smaller
$A_0/M_{1/2}$'s, the $\Dhp=0$ line keeps moving downwards and gets
away from most of the otherwise allowed part of the $\Dst=0$ line.
Also, the intersection of these two lines moves to higher values
of $M_{1/2}$ and $m_0$ and the effect of the $H$-pole is weakened
even around this intersection. As a consequence, the overall
allowed area quickly disappears as $A_0/M_{1/2}$ moves below
$-1.6$, as we will see in \Sref{results}.

\begin{figure}[!t]
\vspace*{-.35in}
\centerline{\epsfig{file=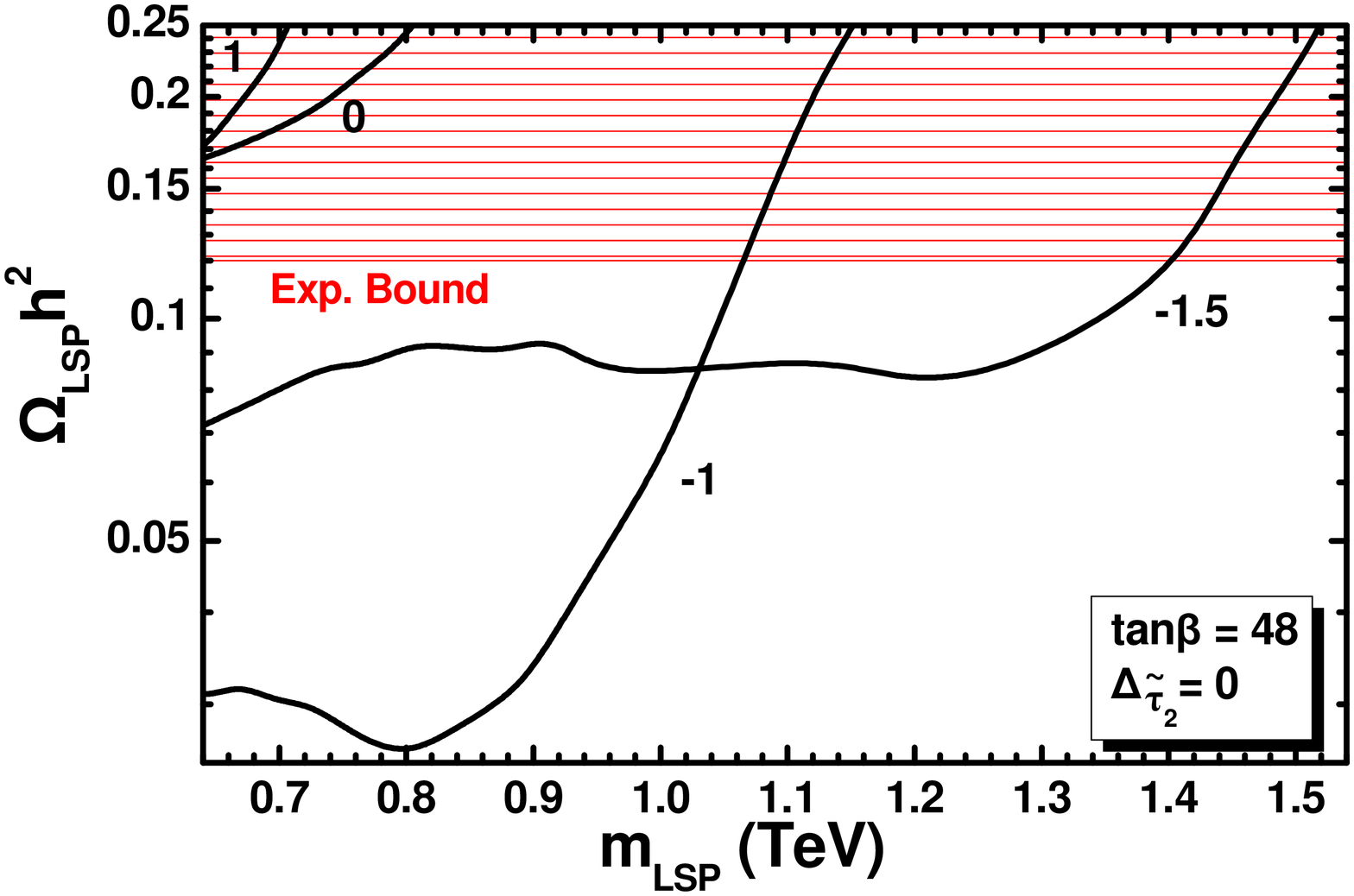,angle=-90,width=11cm}}
\hfill \caption{\sl\ftn $\Omx$ as a function of $m_{\rm LSP}$ for
$\tnb=48,~\Dst\simeq0$, and various $\AMg$'s indicated in the
graph. The experimentally excluded (red) area is also depicted.}
\label{Omf}
\end{figure}

The effect of the $H$-pole on $\Omx$ can be further highlighted by
considering \Fref{Omf}, where we depict $\Omx$ as a function of
$m_{\rm LSP}$ for $\tnb=48,~\Dst\simeq0$, and $\AMg=1,~0,~-1$, and
$-1.5$. We notice that, for $\AMg\geq0$, $\Omx$ is always greater
than $0.12$ and increases sharply with $\mx$. On the
contrary, for $\AMg<0$, $\Omx$ can be smaller than $0.12$ with
an almost flat plateau. More precisely, we see that $\Omx$ remains
almost constant and lower than $0.12$ when $\mx$ is lower than its
value $m_{\rm LSP}^{\rm c}$ at which the $\Dhp=0$ line intersects
the $\Dst=0$ line -- recall that $\mx\simeq0.45\Mg$. From our code,
we estimate that $m_{\rm LSP}^{\rm c}\simeq0.8~\TeV$ [$m_{\rm LSP}^
{\rm c}\simeq1.25~\TeV$] for $\AMg=-1$ [$\AMg=-1.5$]. At $\mx=m_
{\rm LSP}^{\rm c}$, we get a mild temporary reduction of $\Omx$,
whereas, for $\mx>m_{\rm LSP}^{\rm c}$, $\Omx$ increases sharply.

\section{Restrictions on the Supersymmetry Parameters}
\label{results}

\begin{figure}[!t]
\vspace*{-.35in}
\centerline{\epsfig{file=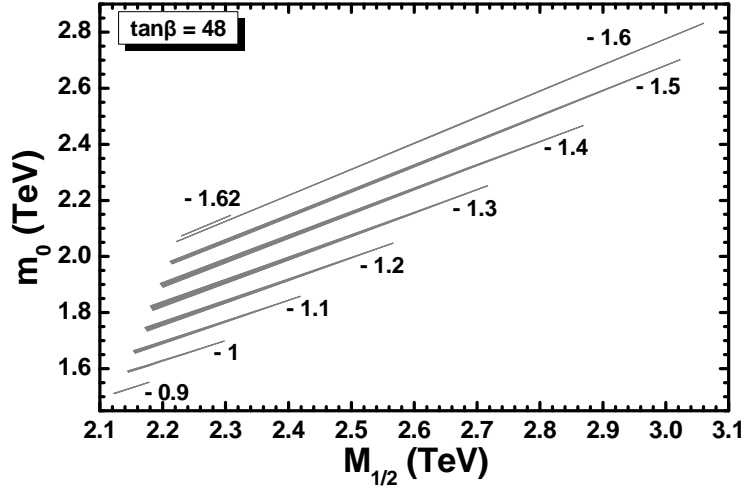,angle=-90,width=11cm}}
\caption{\sl\ftn  The allowed (shaded) areas in the $\Mg-m_0$
plane for $\tan\beta=48$ and various $A_0/\Mg$'s indicated in the
graph.} \label{fig:tanb48m12m0}
\end{figure}

Imposing the requirements described in Secs.~\ref{sec:pheno} and
\ref{sec:cosmo}, we can delineate the allowed para\-meter space of
our model. We find that the only constraints which play a role
are the CDM bound in Eq.~(\ref{cdmb}), the lower bound on $m_h$ in
Eq.~(\ref{mhb}), and the bound on $\bmm$ in Eq.~(\ref{bmmb}). In
the parameter space allowed by these requirements, all the other
restrictions of Sec.~\ref{sec:pheno} are automatically satisfied
with the exception of the lower bound on $\delta a_\mu$ in
Eq.~(\ref{g2b}). This bound is not imposed here as a strict
constraint on the parameters of the model for the reasons
explained in Sec.~\ref{phenod}. We only discuss at which level
\Eref{g2b1} is satisfied in the parameter space allowed by all
the other requirements.

Initially, we concentrate on a representative value of
$\tan\beta=48$ and delineate the allowed areas in the
$M_{1/2}-m_0$ plane for various values of $A_0/\Mg$. These allowed
areas are the shaded ones in Fig.~\ref{fig:tanb48m12m0}. We
observe that these areas are very thin strips. Their lower
boundary corresponds to $\Dst=0$. The area below this boundary is
excluded because the LSP is the charged $\tilde\tau_2$. The upper
boundary of the areas comes from the CDM bound in
Eq.~(\ref{cdmb}), while the left one originates from the limit on
$\bmm$ in Eq.~(\ref{bmmb}). The upper right corner of the areas
coincides with the intersection of the lines $\Dst=0$ and
$\Omx=0.12$. We observe that the allowed area, starting from being
just a point at a value of $A_0/\Mg$ slightly bigger than $-0.9$,
gradually expands as $A_0/\Mg$ decreases and reaches its maximal
size around $A_0/\Mg=-1.6$. For smaller $A_0/\Mg$'s, it shrinks
very quickly and disappears just after $A_0/\Mg=-1.62$. We find
that, for $\tan\beta=48$, $m_{\rm LSP}$ ranges from about 983 to
$1433~{\rm GeV}$, while $m_h$ ranges from about 123.7 to
$125.93~{\rm GeV}$.

\begin{figure}[!t]
\vspace*{-.35in}
\centerline{\epsfig{file=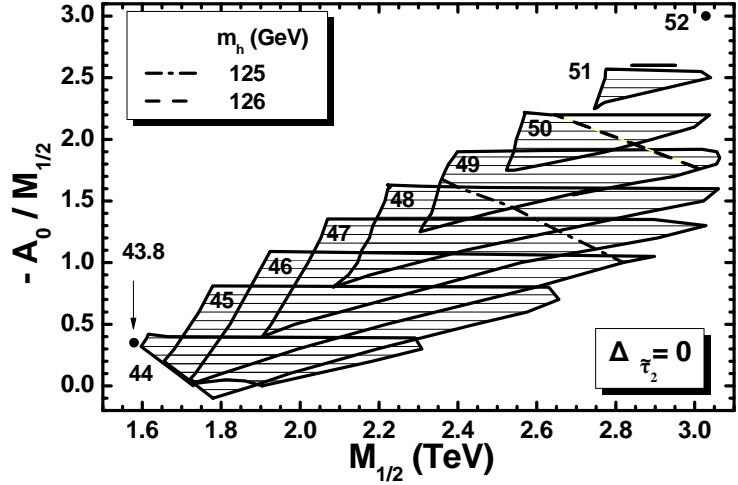,angle=-90,width=11cm}}
\caption{\sl\ftn Allowed regions in the $M_{1/2}-A_{0}/M_{1/2}$
plane for $\Delta_{\tilde\tau_2}=0$ and various $\tnb$'s indicated
in the graph. The dot-dashed [dashed] line corresponds to $m_h=125$
$[126]~{\rm GeV}$.} \label{fig:AMgx}
\end{figure}

\begin{figure}[!t]
\centerline{\epsfig{file=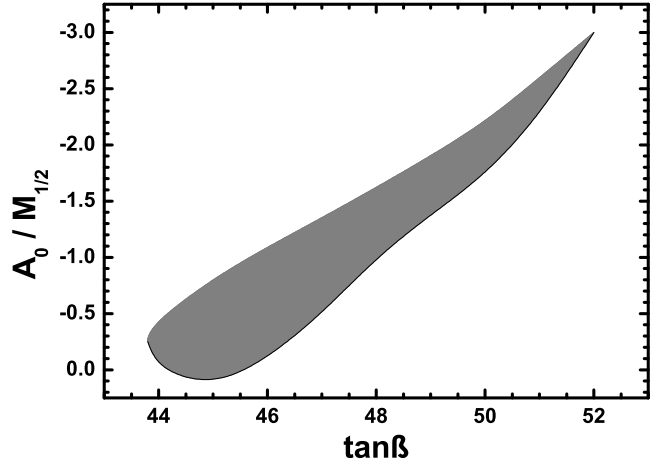,angle=-0,width=9cm}}
\caption{\sl\ftn  The overall (shaded) allowed parameter space of
the model in the $tan\beta-A_0/\Mg$ plane.} \label{fig:A0tanb}
\end{figure}

To get a better idea of the allowed parameter space, we focus on
the coannihilation regime and construct the allowed region in the
$\Mg-A_0/\Mg$ plane. This is shown in \Fref{fig:AMgx}, where we
depict the (horizontally hatched) areas allowed by the various
constraints for $\Dst=0$ and various values of $\tan\beta$ indicated
in the graph. This choice ensures the maximal possible reduction of
$\Omx$ due to the $\tilde\chi\tilde\tau_2$ coannihilation. So, for
$\Dst=0$, we
find the maximal $\Mg$ or $\mx$ allowed by \Eref{cdmb} for a given
value of $A_0/\Mg$. The right boundaries of the allowed regions
correspond to $\Omx=0.12$, while the left ones saturate the bound
on $\bmm$ in Eq.~(\ref{bmmb}) -- cf. \Fref{fig:A0tanb}. The almost
horizontal upper boundaries correspond to the sudden shrinking of
the allowed areas which, as already discussed, is due to the
weakening of the $H$-pole effect as $A_0/M_{1/2}$ drops below a
certain value for each $\tan\beta$. The lower left boundary of the
areas for $\tan\beta=44$, 45, and 46 comes for the lower bound on
$m_h$ in Eq.~(\ref{mhb}), while the somewhat curved, almost
horizontal, part of the lower boundary of the area for
$\tan\beta=44$ originates from the CDM bound in Eq.~(\ref{cdmb}).
The dot-dashed and dashed lines correspond to $m_h=125$ and
$126~{\rm GeV}$ respectively. We see that the $m_h$'s which are
favored by LHC can be readily obtained in our model for the higher
allowed values of $\tan\beta$.

The overall allowed parameter space can be designed in the
$tan\beta-A_0/\Mg$ plane as shown in Fig.~\ref{fig:A0tanb}. Each
point in the shaded space of this figure corresponds to an allowed
area in the $M_{1/2}-m_0$ plane similar to the thin strips shown
in \Fref{fig:tanb48m12m0}. The lower boundary of the allowed
parameter space in Fig.~\ref{fig:A0tanb} originates from the limit
on $\bmm$ in Eq.~(\ref{bmmb}), except its leftmost part which
comes from the lower bound on $m_h$ in Eq.~(\ref{mhb}) or the CDM
bound in Eq.~(\ref{cdmb}). The upper boundary comes from the CDM
bound in Eq.~(\ref{cdmb}). We see that $\tan\beta$ ranges from
about 43.8 to 52. These values are only a little smaller than the
ones obtained in the case of exact YU or the monoparametric YQUCs
discussed in Refs.~\citen{qcdm,nova,nova2,yqu,pekino, pekinoa}. This
mild reduction of $\tan\beta$ is, however, adequate to reduce the
extracted $\bmm$ to an acceptable level compatible with the CDM
requirement. In the allowed area of Fig.~\ref{fig:A0tanb}, the
parameter $A_0/\Mg$ ranges from about $-3$ to 0.1. We also find
that, in this allowed area, the Higgs mass $m_h$ ranges from 122
to $127.23~{\rm GeV}$ and the LSP mass $m_{\rm LSP}$ from about
746.5 to $1433~{\rm GeV}$. So we see that, although $m_h$'s
favored by LHC can be easily accommodated, the lightest neutralino
mass is large making its direct detection very difficult. At the
maximum allowed $\mx$, $\bmm$ takes its minimal value in the
allowed parameter space. This value turns out to be about
$3.64\times10^{-9}$ and, thus, the lower bound in \Eref{bmmb12} is
satisfied everywhere in the allowed area in Fig.~\ref{fig:A0tanb}.
The range of the discrepancy $\delta a_\mu$ between the measured
muon anomalous magnetic moment and its SM value in the allowed
parameter space of Fig.~\ref{fig:A0tanb} is about $\lf0.35-2.76
\rg\times10^{-10}$ (note that $\delta a_\mu$ decreases as $\tnb$
or $\Mg$ increases). Therefore, \Eref{g2b1} is satisfied only at
the level of 2.55 to $2.82-\sigma$. Note that, had we considered
the $\mu<0$ case, $\delta a_\mu$ would have been negative and the
violation of \Eref{g2b1} would have certainly been stronger than
for $\mu>0$.

In Table~\ref{tab:spectrum}, we list the input and the output
parameters of the present model, the masses in $\TeV$ of the SUSY
particles (gauginos/higgsinos
$\tilde\chi,~\tilde\chi_2^{0},~\tilde{\chi}_{3}^{0},~
\tilde{\chi}_{4}^{0},~\tilde{\chi}_{1}^{\pm},~
\tilde{\chi}_{2}^{\pm},~\tilde{g}$, squarks $\tilde{t}_1,~
\tilde{t}_2,~\tilde{b}_1,~\tilde{b}_2,~\tilde{u}_{L},~
\tilde{u}_{R},~\tilde{d}_{L},~\tilde{d}_{R}$, and sleptons
$\tilde\tau_1,~\tilde\tau_2,~\tilde\nu_\tau,~\tilde{\nu}_{e},~
\tilde{e}_L,~\tilde{e}_R$) and  the Higgses ($h$, $H$, $H^\pm$,
$A$), and the values of the various low energy observables in four
characteristic cases (recall that
$1~\pb\simeq2.6\times10^{-9}~\GeV^{-2}$). Note that we have
considered the squarks and sleptons of the two first generations
as degenerate. From the values of the various observable
quantities, it is easy to verify that all the relevant constraints
are met. In the low energy observables, we included the
spin-independent (SI) and spin-dependent (SD) lightest
neutralino-proton ($\xx p$) scattering cross sections $\ssi$ and
$\ssd$, respectively, using central values for the hadronic inputs
-- for the details of the calculation, see Ref.~\citen{yqu}. We
see that, these cross sections are well below not only the present
experimental upper bounds, but even the projected sensitivity of
all planned future experiments. So, the allowed parameter space of
our model will not be accessible to the planned CDM direct
detection experiments based on neutralino-proton scattering. We
also notice that, the sparticles turn out to be very heavy, which
makes their discovery a very difficult task.

\begin{table}[!t]
\tbl{Input/output parameters, sparticle and Higgs masses, and low
energy observables in four cases. }
{\begin{tabular}{c@{\hspace{0.3cm}}c@{\hspace{0.3cm}}c@
{\hspace{0.3cm}}c@{\hspace{0.3cm}}c} \toprule
\multicolumn{5}{c}{Input Parameters}\\\colrule
$\tan\beta$ & $48$ & $49$ & $50$ & $51$\\
$-A_0/\Mg$ &$1.4$ &$1.6$&$2$ &$2.5$ \\
$\Mg/\TeV$ & $2.27$ &$2.411$&$2.824$ &$2.808$ \\
$m_0/\TeV$ &$1.92$ &$2.295$&$3.156$&$3.747$ \\ \colrule
\multicolumn{5}{c}{Output Parameters}\\\colrule
$h_t/h_\tau(M_{\rm GUT})$ & $1.117$ & $1.079$ & $1.038$ & $1.008$\\
$h_b/h_\tau(M_{\rm GUT})$ & $0.623$ & $0.618$ & $0.613$ & $0.607$\\
$h_t/h_b(M_{\rm GUT})$ & $1.792$ & $1.745$ & $1.693$ &
$1.660$\\
$\mu/\TeV$ & $2.78$ & $3.092$ & $3.823$ & $4.129$\\
$\Dst (\%)$ & $1.43$ & $0.93$ & $0.1$ & $0.17$\\
$\Delta_H (\%)$ & $3.08$ & $1.30$ & $0.11$ & $1.76$\\
\colrule
\multicolumn{5}{c}{Masses in ${\rm TeV}$ of Sparticles and
Higgses}\\\colrule
$\tilde\chi, \tilde\chi_2^{0}$& $1.023, 1.952$ &$1.110, 2.117$ &$1.309, 2.489$ &$1.303, 2.481$\\
$\tilde{\chi}_{3}^{0}, \tilde{\chi}_{4}^{0}$ &$2.782, 2.785$ &$3.088, 3.091$ &$3.815, 3.817$ &$4.114, 4.116$\\
$\tilde{\chi}_{1}^{\pm}, \tilde{\chi}_{2}^{\pm}$ &$1.985, 2.785$ &$2.117, 3.091$ &$2.489, 3.817$ &$2.481, 4.116$\\
$\tilde{g}$&$4.809$ &$5.190$ &$6.042$ &$6.040$ \\ 
%
$\tilde{t}_1, \tilde{t}_2$ &$3.806, 3.226$ &$4.097, 3.458$ &$4.761, 3.967$ &$4.781, 3.902$\\
$\tilde{b}_1, \tilde{b}_2$ &$3.838, 3.763$ &$4.141, 4.058$ &$4.853, 4.733$ &$4.947, 4.757$ \\
$\tilde{u}_{L}, \tilde{u}_{R}$&$4.687, 4.485$ &$5.138, 4.923$ &$6.186, 5.946$ &$6.483, 6.257$\\
$\tilde{d}_{L}, \tilde{d}_{R}$& $4.687, 4.459$ &$5.138, 4.896$ &$6.187, 5.914$ &$6.483, 6.227$\\
$\tilde\tau_1, \tilde\tau_2$&$2.082, 1.037$ &$2.347, 1.121$ &$2.979, 1.310$ &$3.293, 1.305$\\
$\tilde\nu_\tau, \tilde{\nu}_{e}$ &$2.075, 2.451$ &$2.342, 2.818$ &$2.975, 3.689$ &$3.289, 4.200$\\
$\tilde{e}_L, \tilde{e}_R$&$2.453, 2.112$ &$2.819, 2.476$ &$3.690, 3.339$ &$4.201, 3.901$\\
\colrule
%
$h, H$&$0.1245, 2.109$ &$0.125, 2.249$ &$0.126, 2.621$&$0.1265, 2.652$\\
$H^{\pm}, A$&$2.111, 2.110$  &$2.251, 2.25$ &$2.623,
2.622$&$2.654, 2.652$ \\\colrule
\multicolumn{5}{c}{Low Energy Observables}\\\colrule
%
$10^4\bsg$ &$3.25$  &$3.25$ &$3.26$&$3.26$\\
$10^9\bmm$   &$4.17$ &$4.15$ &$3.98$&$4.17$\\
$\btn$   &$0.975$  &$0.977$ &$0.982$&$0.982$\\
$10^{10}\Dam$  &$1.11$&$0.89$ &$0.57$ &$0.49$\\\colrule $\Omx$
&$0.11$&$0.11$&$0.11$&$0.11$\\\colrule
$\ssi / 10^{-12} \pb$ &$6.17$&$4.55$ &$2.44$&$1.75$\\
$\ssd / 10^{-9} \pb$
&$1.69$&$1.08$ &$0.43$&$0.28$\\
\botrule
\end{tabular}\label{tab:spectrum}}
\end{table}

The fact that, in our model, $\Mg$, $m_0$, and $\mu$ generally
turn out to be of the order of a few $\TeV$ puts under some stress
the naturalness of the radiative EWSB -- see \Eref{mu}. This is,
though, a general problem of the CMSSM, especially in view of the
recent data on $B_s\to\mu^+\mu^-$ and $m_h$ as noted in
\cref{CmssmLhc, CmssmLhca, CmssmLhcb}. Attempts to address this
problem, known as little hierarchy problem, invoke departures from
the CMSSM universality \cite{Min, natural, naturala, naturalb,
naturalc} or addition of extra matter superfields \cite{Hall} --
singlet \cite{king, kinga, kingb} or vector-like \cite{martin,martin2, martina} --
beyond the MSSM ones.

\section{The Deviation from Yukawa Unification} \label{viol}

\begin{figure}[!t]
\centering 
\begin{minipage}{\textwidth}
\epsfig{file=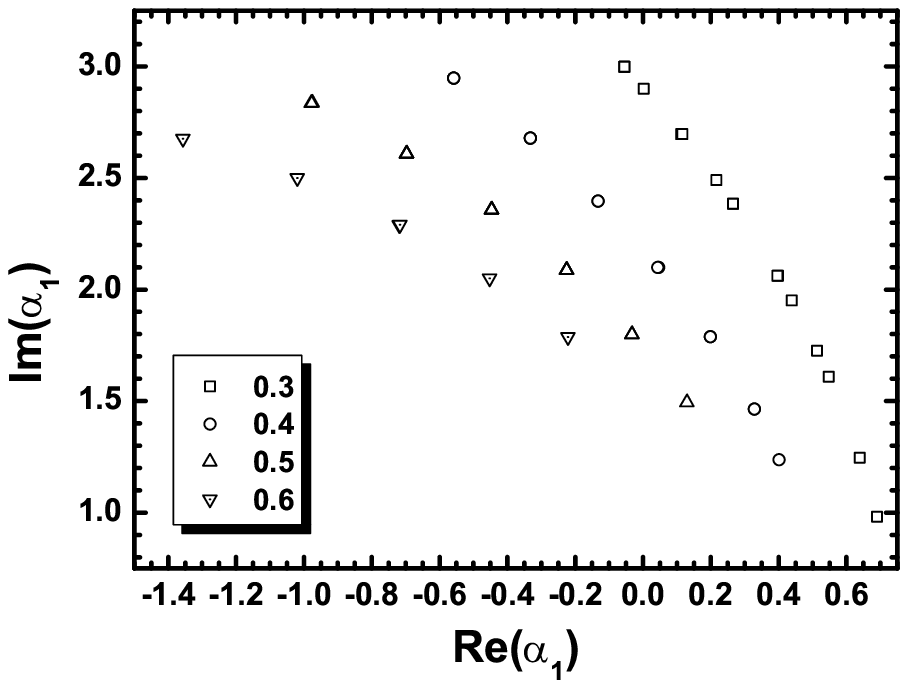,height=1.9in,angle=-0}
\epsfig{file=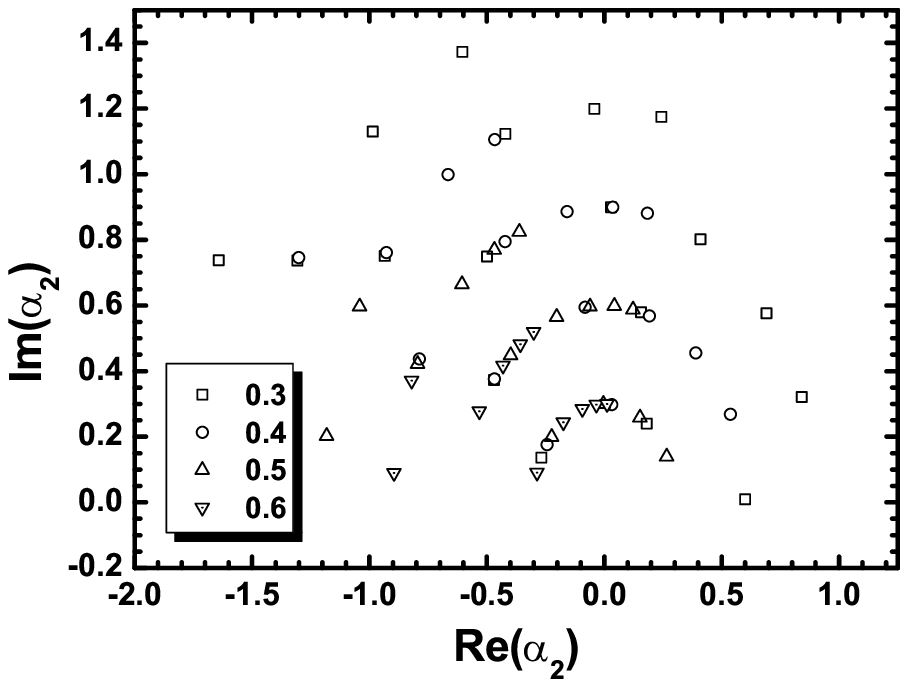,height=1.9in,angle=-0} 
\end{minipage}
\caption{\sl\ftn The complex parameters $\alpha_1$ and $\alpha_2$
for various values of the real and positive parameter $\rho$,
which are indicated in the graphs, for the case presented in the
second column of Table~\ref{tab:spectrum} with $tan\beta=49$ and
$m_h=125~{\rm GeV}$.} \label{fig:a1a2}
\end{figure}

In the overall allowed parameter space of our model in
Fig.~\ref{fig:A0tanb}, we find the following ranges for the ratios
of the asymptotic third generation Yukawa coupling constants:
$h_t/h_\tau\simeq 0.98-1.29$, $h_b/h_\tau\simeq 0.60-0.65$, and
$h_t/h_b\simeq 1.62-2.00$. We observe that, although exact YU is
broken, these ratios remain close to unity. They can generally be
obtained by natural values of the real and positive parameter
$\rho$ and the complex parameters $\alpha_1$, $\alpha_2$, which
enter the YQUCs in Eq.~(\ref{quasi}). Comparing these ratios with
the ones of the gauge coupling constants of the non-SUSY SM at a
scale close to $\Mgut$ -- see e.g. \cref{book} --, we can infer that
the ratios here are not as close to unity. Despite this fact, we
still apply the term `Yukawa quasi-unification' in the sense that the
ratios of the Yukawa coupling constants in our model are much closer
to unity than in generic models with lower values of $\tan\beta$ -- cf.
\cref{Antuch,Antuch2}. Finally, note that the deviation from exact YU here
is comparable to the one obtained in the monoparametric case --
cf. \cref{yqu} -- and is also generated in a natural, systematic,
controlled, and well-motivated manner.

In order to see that these ratios can be obtained by natural values
of $\rho$, $\alpha_1$, and $\alpha_2$, we take as a characteristic
example the second out of the four cases presented in
Table~\ref{tab:spectrum}, which yields $m_h=125~{\rm GeV}$ favored
by the LHC. In this case, where $h_b/h_\tau=0.618$ and
$h_t/h_\tau=1.079$, we solve Eq.~(\ref{quasi}) w.r.t. the complex
parameters $\alpha_1$, $\alpha_2$ for various values of the real
and positive parameter $\rho$. Needless to say that one can find
infinitely many solutions, since we have only two equations and
five real unknowns. Some of these solutions are shown
Fig.~\ref{fig:a1a2}. Note that the equation for $h_b/h_\tau$
depends only on the combination $\rho\alpha_1$ and, therefore, its
solutions are expected to lie on a certain curve in the complex
plane of this combination. Consequently, in the $\alpha_1$ complex
plane, the solutions should be distributed on a set of similar
curves corresponding to the various values of the real and positive
parameter $\rho$. This is indeed the case as one can see from the
left panel of Fig.~\ref{fig:a1a2}. For each $\alpha_1$ and $\rho$
in this panel, we then solve the equation for $h_t/h_\tau$ to find
the complex parameter $\alpha_2$. In the right panel of
Fig.~\ref{fig:a1a2}, we show
several such solutions. Observe that the equation for $h_t/h_\tau$
depends separately on $\alpha_2$ and $\rho$ and, thus, its solutions
do not follow any specific pattern in the $\alpha_2$ complex plane.
Note that each point in the $\alpha_1$ complex plane generally
corresponds to more than one points in the $\alpha_2$ complex plane.
We scanned the range of $\rho$ from 0.3 to 3 and found solutions
only for the lower values of this parameter (up to about 0.6). The
solutions found for $\alpha_1$ and $\alpha_2$ are also limited in
certain natural regions of the corresponding complex planes. The
picture is very similar to the one just described for all the
possible values of the ratios of the third generation Yukawa
coupling constants encountered in our investigation. So, we
conclude that these ratios can be readily obtained by a multitude
of natural choices of the parameters $\rho$, $\alpha_1$, and
$\alpha_2$ everywhere in the overall allowed parameter space of
the model.

\section{The Hyperbolic Branch/Focus Point Area}
\label{hbfp}

It is generally accepted that the mSUGRA/CMSSM parameter space has
been significantly squeezed by the recent experimental results. In
particular, in most of the allowed parameter space, the LSP and
the other sparticles turn out to be too heavy and a mild tuning is
required for achieving the radiative EWSB. As discussed above, the
CMSSM with (generalized) Yukawa quasi-unification is not an
exception at least in the $H$-pole enhanced stau-antistau
coannihilation regime considered here. There exists, though, a
broader viable region of the parameter space of the CMSSM, which
does not require excessive electroweak fine tuning and can yield a
relatively light LSP. This region, which is characterized by large
values of the ratio $m_0/M_{1/2}$, is known as the HB/FP regime
\cite{focus, focusa, focusb, focusc, focusd, focuse, focusf}.

In our model, we also find a viable HB/FP area with the LSP still
being the lightest neutralino, but now with a significant higgsino
component. Its mass can even be well below $100~\GeV$. This
area extends to large $m_0$'s ($>15~\TeV$) and $M_{1/2}$'s, but its
most interesting part is the one at low $M_{1/2}$'s. In this part
and for small $m_0$'s, the reduction of
$\Omx$ is mostly caused by $\xx\xx$ annihilation effects. As $m_0$
gets larger, the neuralino-chargino and chargino-chargino coannihilation
effects become dominant. The constraints from $B$ physics are all well
satisfied in this regime. On the other hand, the value of $\delta a_\mu$
is still below the lower bound in Eq.~(\ref{g2b}). The ratios of the
third generation Yukawa coupling constants remain close to unity.
Having in mind the latest indications for a light candidate CDM particle
from the CDMS II experiment \cite{cdmsii}, we see that the HB/FP region
is very promising and it definitely needs more light to be shed on it.
We are currently pursuing the detailed investigation of this region.

\section{Conclusions} \label{con}

We performed an analysis of the parameter space of the CMSSM with
$\mu>0$ supplemented by a generalized asymptotic Yukawa coupling
quasi-unification condition, which is implied by the SUSY GUT
constructed in Ref.~\citen{qcdm} and allows an experimentally
acceptable $b$-quark mass. We imposed a number of cosmological and
phenomenological constraints originating from the CDM abundance in
the universe, $B$ physics ($b \rightarrow s\gamma$, $B_s\to
\mu^+\mu^-$, and $B_u\to\tau\nu$), and the mass $m_h$ of the
lightest neutral CP-even Higgs boson. We found that the lightest
neutralino can act as a CDM candidate in a relatively wide range
of parameters. In particular, the upper bound from CDM
considerations on the lightest neutralino relic abundance, which
is drastically reduced mainly by $H$-pole enhanced stau-antistau
coannihilation processes, is compatible with the recent data on
the branching ratio of $B_s\to\mu^+\mu^-$ in this range of
parameters. Also, values of $m_h\simeq(125-126)~{\rm GeV}$, which
are favored by the LHC, can be easily accommodated. The mass of
the lightest neutralino, though, comes out to be large ($\sim
1~{\rm TeV}$), which makes its direct detectability very difficult
and the sparticle spectrum very heavy.

\section*{Acknowledgments}

This work was supported by the European Union under the Marie
Curie Initial Training Network `UNILHC' PITN-GA-2009-237920 and
the Greek Ministry of Education, Lifelong Learning, and Religious
Affairs and the Operational Program: Education and Lifelong
Learning `HERACLITOS II'.

\def\ijmp#1#2#3{{\it Int. Jour. Mod. Phys.}
{\bf #1},~#3~(#2)}
\def\plb#1#2#3{{\it Phys. Lett. B }{\bf #1},~#3~(#2)}
\def\zpc#1#2#3{{\it Z. Phys. C }{\bf #1},~#3~(#2)}
\def\prl#1#2#3{{\it Phys. Rev. Lett.}
{\bf #1},~#3~(#2)}
\def\rmp#1#2#3{{\it Rev. Mod. Phys.}
{\bf #1},~#3~(#2)}
\def\prep#1#2#3{{\it Phys. Rep. }{\bf #1},~#3~(#2)}
\def\prd#1#2#3{{\it Phys. Rev. D }{\bf #1},~#3~(#2)}
\def\npb#1#2#3{{\it Nucl. Phys. }{\bf B#1},~#3~(#2)}
\def\npps#1#2#3{{\it Nucl. Phys. B (Proc. Sup.)}
{\bf #1},~#3~(#2)}
\def\mpl#1#2#3{{\it Mod. Phys. Lett.}
{\bf #1},~#3~(#2)}
\def\arnps#1#2#3{{\it Annu. Rev. Nucl. Part. Sci.}
{\bf #1},~#3~(#2)}
\def\sjnp#1#2#3{{\it Sov. J. Nucl. Phys.}
{\bf #1},~#3~(#2)}
\def\jetp#1#2#3{{\it JETP Lett. }{\bf #1},~#3~(#2)}
\def\app#1#2#3{{\it Acta Phys. Polon.}
{\bf #1},~#3~(#2)}
\def\rnc#1#2#3{{\it Riv. Nuovo Cim.}
{\bf #1},~#3~(#2)}
\def\ap#1#2#3{{\it Ann. Phys. }{\bf #1},~#3~(#2)}
\def\ptp#1#2#3{{\it Prog. Theor. Phys.}
{\bf #1},~#3~(#2)}
\def\apjl#1#2#3{{\it Astrophys. J. Lett.}
{\bf #1},~#3~(#2)}
\def\n#1#2#3{{\it Nature }{\bf #1},~#3~(#2)}
\def\apj#1#2#3{{\it Astrophys. J.}
{\bf #1},~#3~(#2)}
\def\anj#1#2#3{{\it Astron. J. }{\bf #1},~#3~(#2)}
\def\mnras#1#2#3{{\it MNRAS }{\bf #1},~#3~(#2)}
\def\grg#1#2#3{{\it Gen. Rel. Grav.}
{\bf #1},~#3~(#2)}
\def\s#1#2#3{{\it Science }{\bf #1},~#3~(#2)}
\def\baas#1#2#3{{\it Bull. Am. Astron. Soc.}
{\bf #1},~#3~(#2)}
\def\ibid#1#2#3{{\it \it ibid. }{\bf #1},~#3~(#2)}
\def\cpc#1#2#3{{\it Comput. Phys. Commun.}
{\bf #1},~#3~(#2)}
\def\astp#1#2#3{{\it Astropart. Phys.}
{\bf #1},~#3~(#2)}
\def\epjc#1#2#3{{\it Eur. Phys. J. C}
{\bf #1},~#3~(#2)}
\def\nima#1#2#3{{\it Nucl. Instrum. Meth. A}
{\bf #1},~#3~(#2)}
\def\jhep#1#2#3{{\it J. High Energy Phys.}
{\bf #1},~#3~(#2)}
\def\jcap#1#2#3{{\it J. Cosmol. Astropart. Phys.}
{\bf #1},~#3~(#2)}
\def\apjs#1#2#3{{\it Astrophys. J. Suppl.}
{\bf #1},~#3~(#2)}

\end{document}